\documentclass[journal]{IEEEtran}

\usepackage{cite}
\usepackage{amsmath,amssymb,amsfonts}
\usepackage{algorithmic}
\usepackage{graphicx}
\usepackage{textcomp}
\usepackage{xcolor}
\usepackage{bm}
\usepackage[caption=false]{subfig}
\usepackage{comment}
\usepackage{float}
\usepackage{svg}
\usepackage{color, soul}
\usepackage{balance}
\usepackage{siunitx}
\usepackage{dsfont}
\usepackage{enumitem}

\usepackage[acronym,shortcuts]{glossaries}
\newacronym{awgn}{AWGN}{additive white Gaussian noise}

\newacronym{cdf}{CDF}{cumulative distribution function}
\newacronym{cr}{CR}{challenge-response}

\newacronym{det}{DET}{detection error trade-off}

\newacronym{fa}{FA}{false alarm}

\newacronym{gat}{GAT}{gain authentication test}
\newacronym{glrt}{GLRT}{generalized likelihood ratio test}

\newacronym{llt}{LLT}{log-likelihood test}
\newacronym{lrt}{LRT}{likelihood ratio test}

\newacronym{md}{MD}{missed detection}
\newacronym{mimo}{MIMO}{multiple-input and multiple-output}
\newacronym{ml}{ML}{maximum likelihood}
\newacronym{mne}{MNE}{multiple Nash equilibrium strategy}

\newacronym[longplural=Nash equilibria]{ne}{NE}{Nash equilibrium}

\newacronym{pla}{PLA}{physical layer authentication}
\newacronym{pdf}{PDF}{probability density function}
\newacronym{pfa}{PFA}{probability of false alarm}
\newacronym{pmd}{PMD}{probability mass distribution}
\newacronym{psk}{PSK}{phase-shift keying}

\newacronym{sne}{S-NE}{single NE}
\newacronym{shm}{SHM}{single-hop multiple}
\newacronym{mrm}{MRM}{multi-round multiple}
\newacronym{nlos}{NLOS}{not line-of-sight}
\newglossaryentry{me_b}{%
name=\ensuremath{\bm{m}_{\rm E}},
description={wavelength}
}

\newglossaryentry{m}{%
name=\ensuremath{\bm{m}},
description={m}
}

\newtheorem{theorem}{Theorem}
\newtheorem{lemma}{Lemma}

\usepackage{pgfplots}
\pgfplotsset{compat=newest}
\pgfplotsset{plot coordinates/math parser=false}
\newlength\fwidth
\newlength\fheight
\DeclareMathOperator*{\argmin}{arg\,min} 
\DeclareMathOperator*{\argmax}{arg\,max} 

\pgfplotsset{ tick label style={font=\scriptsize}, label style={font=\scriptsize}, legend style={font=\scriptsize},
}

\makeatletter
\newcommand{\new}[1]{{\textcolor{blue}{#1}}}

\makeatletter
\newcommand\remembertext[2]{
  \immediate\write\@auxout{\unexpanded{\global\long\@namedef{mytext@#1}{#2}}}%
  {\color{black} #2}%
}
\newcommand\recalltext[1]{%
  \new{\ifcsname mytext@#1\endcsname
    \fontsize{10.5}{12.5}\selectfont\@nameuse{mytext@#1}%
  \else
    ``??''
  \fi
}}
\makeatother

\title{Challenge-Response to Authenticate Drone Communications: A Game Theoretic Approach} 

 \author{ Mattia~Piana,~\IEEEmembership{Student~Member,~IEEE}, \IEEEauthorblockN{Francesco~Ardizzon,~\IEEEmembership{Member,~IEEE}, and Stefano~Tomasin,~\IEEEmembership{Senior~Member,~IEEE}
\thanks{Manuscript received --; revised -- and -- accepted --. Date of publication --; date of current version --.} 
\thanks{Corresponding author: M. Piana. This work was partially supported by the European Union under the Italian National Recovery and Resilience Plan (PNRR) of NextGenerationEU, partnerships on “Telecommunications of the Future” (PE0000001 - program “RESTART and PE00000014 - program "SERICS"), and through the Horizon Europe/JU SNS project ROBUST-6G (Grant Agreement no. 101139068). The authors are with the Department of Information Engineering, Universit\`a degli Studi di Padova, Padua 35131, Italy. S. Tomasin is also with the National Inter-University Consortium for Telecommunications (CNIT), 43124 Parma, Italy. (email:  {\{mattia.piana@phd., francesco.ardizzon@, stefano.tomasin@\}unipd.it}).}
}}

\IEEEoverridecommandlockouts 

\begin{document}
\sloppy
\maketitle

\begin{abstract} \color{black}
As drones are increasingly used in various civilian applications, the security of drone communications is a growing concern. In this context, we propose novel strategies for challenge-response physical layer authentication (CR-PLA) of drone messages. The ground receiver (verifier) requests the drone to move to a defined position (challenge), and authenticity is verified by checking whether the corresponding measured channel gain (response) matches the expected statistic.
In particular, the challenge is derived from a mixed strategy obtained by solving a zero-sum game against the intruder, which in turn decides its own positions. In addition, we derive the optimal strategy for multi-round authentication, where the CR-PLA procedure is iterated over several rounds. We also consider the energy minimization problem, where legitimate users want to minimize the energy consumption without compromising the security performance of the protocol.
The performance of the proposed scheme is tested in terms of both security and energy consumption through numerical simulations, considering different protocol parameters, different scenarios (urban and rural), different drone altitudes, and also in the context of drone swarms.
\end{abstract}



\glsresetall

\section{Introduction}\label{sec:intro}
\IEEEPARstart{T}{he} use of drones has rapidly increased over the last few years. Starting as a military tool, they are now used in many civil applications such as precision agriculture, environmental monitoring, and disaster management and relief. As drones become integrated into more complex systems, security is a rising concern \cite{Adil23systematic}. Among the major threats, the transmission of jamming or spoofing signals is particularly dangerous as these can disrupt the navigation system \cite{ceccato21,michieletto22} and jeopardize the drone mission.

This paper addresses the problem of authenticating messages transmitted by drones. In particular, a drone or ground device (Bob) wants to authenticate messages potentially transmitted by a legitimate drone (Alice) and distinguish them from those transmitted by an {\em intruder} drone (Trudy) attempting to impersonate Alice. Cryptography-based authentication solutions have several drawbacks: a) they are typically computationally expensive, b) they often provide only computational security, which may be vulnerable to quantum computing-based algorithms, and c) they introduce significant communication overhead. 
Both communication overhead and computational complexity lead to high energy consumption, which is a limited resource for drones. Therefore, in this paper, we focus on \ac{pla} mechanisms that provide security by exploiting the statistical properties of the channels. These techniques typically consume less energy than their crypto-based counterparts while providing information-theoretic security.

\ac{pla} has recently been studied in \cite{Illi2024Physical,Hoang2024Physical}, and mechanisms specifically targeting drone communication may rely on fingerprinting \cite{Maeng2021Power}, or on channel characteristics, eventually supported by a federated learning architecture \cite{Yazdinejad2021Federated}. Here we focus on a recent evolution of \ac{pla} using a \ac{cr} approach. In cryptographic \ac{cr} authentication \cite[Sec. 13.5]{gupta2014cryptography}, Alice and Bob share a secret key. Then, Bob sends a message called a \emph{challenge} over a public channel, and Alice computes and sends back to Bob a \emph{response} obtained using a hash function of both the secret key and the challenge. Finally, Bob verifies the authenticity of the sender of the response by comparing the response to a local response obtained using the same secret key and challenge. Instead of relying on crypto-based solutions, we consider \ac{cr}-\ac{pla}, first introduced in \cite{Tomasin22Challenge}. \Ac{cr}-\ac{pla} is based on \emph{partially controllable channels}, where the challenge is issued by the verifier Bob by manipulating the propagation environment, while the response is the received signal from Alice, which must be consistent with the expected change. Since the challenge, i.e., the propagation conditions, is randomly chosen by the verifier, it is difficult for Trudy to predict it and transmit a signal that is consistent with it.

In \cite{Mazzo23Physical} and \cite{ardizzon2024energy}, the authors proposed a first \ac{cr}-\ac{pla} protocol for drone communication. However, a simple channel model was considered, and the challenge distribution was not optimized but chosen as uniform. Additionally, not even the attacker was allowed to optimize their attacks. In this work, we instead consider a more complete channel model and exploit its properties, in particular shadowing, to design the challenge distribution and investigate more sophisticated attacks. 

Apart from our previous work \cite{ardizzon2024energy}, while there are studies on the energy consumption of drones, e.g., \cite{Abeywickrama18}, few studies analyze the link between energy consumption and physical-layer security protocols, as the analyses typically focus on the power consumption of crypto/higher-level solutions, e.g., \cite{Zhang2020Lightweight}.
 
In this paper, we propose novel strategies for Bob and Trudy when using \ac{cr}-\ac{pla} in drone communication. In particular, in a preliminary phase, Bob measures the channel gain when Alice is at a set of predefined positions. Then, the authenticated transmission protocol requires that Bob first randomly selects a set of positions (with a suitable distribution) and secretly communicates them to Alice. Alice goes to the indicated positions and, for each of them, she transmits a pilot signal together with the message to be authenticated. Next, Bob assesses the authenticity of the received signal by checking that the measured channel gains correspond to the expected ones, estimated during the preliminary phase. In this context, the challenge is represented by the set of positions, and the corresponding response is the set of channel gains estimated by Bob. In turn, for her attack, Trudy randomly selects (with a suitable distribution) a set of positions from which to transmit the pilot signal and her message. The distributions used by Bob and Trudy to select the position sets are optimized to their advantage by finding the \acp{ne} of a zero-sum game.
In addition, we also consider the problem of minimizing the energy of Alice's movements by proposing both optimal and heuristic strategies to minimize the (average) distance traveled by Alice without sacrificing the security of the protocol.

The contributions of this paper are as follows 
\begin{itemize}
    \item We model the channel between the drone and the receiver, including both the preliminary channel estimation phase and the security protocol. 
    \item We design the statistical distribution of positions generated by Bob and Trudy by modeling the problem as a zero-sum game between legitimate users and Trudy, where the payoff is the \ac{md} probability for a target \ac{fa} probability.
    \item We consider both optimal and low-complexity solutions for optimizing Bob's position selection statistics.  
    \item We test the proposed technique by numerical simulation, based on a realistic model of both Alice-Bob and Trudy-Bob channels, including shadowing effects \textcolor{black}{in urban and rural scenarios}.
\end{itemize}

Several papers have shown the effectiveness of game theory in authentication. In \cite{Xiao2016Channel}, the authors use game theory to design an authentication protocol for \ac{mimo} channels.  In \cite{Ge2024GAZETA}, a game-theoretic authentication protocol for zero-trust fifth-generation (5G) Internet of Things (IoT) networks is proposed. The multi-agent scenario where multiple receivers and spoofers play has been analyzed in \cite{Wu2023Game}.

A game-theoretic \ac{pla} protocol for drone communication was proposed in \cite{Zhou2022Game}, where Bob chooses the detection thresholds while Trudy plays by choosing the attack probability. However, they consider a more restrictive model where the test is performed by assuming that the previously received signal was authentic, and thus the signal under test should match the previous one. 

The rest of the paper is organized as follows. Section~\ref{sec:sysModel} describes the system model in detail. Section~\ref{sec:aut_scheme} gives an overview of the proposed protocol. The \ac{cr}-\ac{pla} problem is further characterized in Section~\ref{sec:game_form}, while Section~\ref{sec:Security optimization} details its solution using game theory. Numerical results are given in section~\ref{sec:results}. Section~\ref{sec:conclusion} draws the conclusions.

\section{System Model}\label{sec:sysModel}
In the considered scenario, Alice is a legitimate drone transmitting messages to Bob, which can be a ground device or, eventually, a second drone. A third drone, Trudy, sends fraudulent messages to Bob. \remembertext{multi drones}{We remark that, although our approach can be applied to multiple drones that collaborate for a joint swarm authentication against multiple coordinated attackers, we mainly focus on a single-drone scenario. Yet, in Section~\ref{sec:results}, multiple drones are considered.} Receiver Bob aims to distinguish messages coming from Alice and Trudy. To this end, we exploit Alice's mobility to design a \ac{cr}-\ac{pla} protocol. In turn, Trudy aims to fool Bob by forging her transmissions so that Bob confuses Trudy's messages as legitimate, i.e., as coming from Alice.

Without loss of generality, we use a Cartesian coordinate system, centered on Bob, to describe any device position. Let $\bm{x} = (x_1, x_2, x_3)$ be the generic position of Alice and $\bm{y} = (y_1, y_2, y_3)$ the generic position of Trudy. 
In particular, we consider the space where Alice and Trudy move to be sampled by a plane grid at altitude $h$ containing $M_\mathrm{A}$ and $M_\mathrm{T}$ positions, respectively. Thus, Alice's positions are the set
\begin{equation} \label{eq:ALice_set_pos}
\mathcal{X} = \left\{\bm{x}(i)=(x_{i,1},x_{i,2},h), \quad i=1,\ldots, M_\mathrm{A}\right\}.
\end{equation}
\remembertext{al_tr_asymmetry}{Similarly to Alice, Trudy can move to positions in a plane grid at the same Alice's altitude $h$, collected into the set} 
\begin{equation}
\mathcal{Y} = \{\bm{y}(j)=(y_{j,1},y_{j,2},h), \quad j=1,\ldots, M_\mathrm{T}\}.
\end{equation}

While considering a finite set of positions (e.g., on a grid) makes the design of the authentication protocol easier, i) the results can be straightforwardly extended to a general 3D model ii) the map can be very dense allowing Alice to go everywhere on the grid, up to the accuracy of the drone positioning system.

Exploiting Alice's mobility for authentication purposes also entails an energy cost. \remembertext{droneEnergy1}{As Alice's aim is minimizing such cost, we consider Alice moving from position $\bm{x}(i)$ to $\bm{x}(i')$ on a straight line. The cost is then denoted as $E(\bm{x}(i),\bm{x}(i'))$.}
In particular, assuming Alice moves at a constant speed, with magnitude $v$, on the horizontal plane, the energy consumption (in Joules) is a linear function of the distance \cite{Abeywickrama18}, namely
\begin{equation}\label{eq:energy}
    E(\bm{x}(i),\bm{x}(i')) = \alpha_1\, \frac{ d_{i,i'}}{v} - \alpha_0\,, 
\end{equation}
where $\alpha_1$ and $\alpha_0$ are constants parameters and the distance between position $\bm{x}(i)$ and $\bm{x}(i')$ is
\begin{equation}\label{eq:def_dist}
    d_{i,i'}=\| \bm{x}(i') -\bm{x}(i)\|.
\end{equation}

Hence, since in \eqref{eq:energy} the energy is a function of the traveled distance, in the following we will consider the traveled distance instead of the energy as a cost metric\footnote{\remembertext{droneEnergy2}{It could also be possible to include the energy spent to reach the maximum velocity. However, we will consider a grid step larger than the distance required to reach such velocity. Thus, the energy consumption due to the initial acceleration is just an additional constant term in \eqref{eq:energy}.}}.

\subsection{Channel Model}
Communications among devices occur through narrowband \ac{awgn} channels, and their attenuations include path loss, shadowing, and fading. Path loss takes into account the average attenuation caused by the distance traveled by the signal, while shadowing accounts for the presence of objects/obstacles in the environment and is a slow-fading phenomenon, i.e., it changes slowly over time.  

Let $A_\mathrm{Tx}$ and $ A_\mathrm{Rx}$ be the gains of the transmitter and receiver antennas, respectively, in dB. Letting $(f_{\rm c})_{\si{\mega\hertz }}$ be the carrier frequency in \si{\mega\hertz} and $(d)_{\si{\kilo\meter }} = \|\bm{q}\|$ the distance (in \si{\kilo\meter}) between transmitter in a generic position $\bm{q}$ and the receiver Bob, we define the path-loss attenuation according to the Friis formula (in \si{\decibel}) as
\begin{equation}\label{eq:pathloss}
\begin{split}
(\xi_{\bm{q}})_{\rm dB}= &32.4+20\log_{10}{(d)_{\si{\kilo\meter }}}+20\log_{10}{(f_{\rm c})_{\si{\mega\hertz }}}  + \\
 &   -A_\mathrm{Tx}- A_\mathrm{Rx}.
\end{split}
\end{equation}

According to the Gudmundson model \cite{gudmundson1991correlation}, the shadowing attenuation in \si{\decibel} $(s)_{\rm dB}$ follows a zero-mean normal distribution with variance $\sigma^{2}_{(s)_{\rm dB}}$, in short $(s)_{\rm dB} \sim \mathcal{N}(0,\,\sigma^{2}_{(s)_{\rm dB}})$. The shadowing attenuations $(s)_{\rm dB}$ and $(s)'_{\rm dB}$, measured at positions $\bm{q}$ and $\bm{q}'$, are correlated: defining $d_\mathrm{ref}$ as the coherence distance \cite{bentom}, the correlation is 
\begin{equation}\label{eq:gudmundson}
\mathbb{E}((s)_{\rm dB}(s)'_{\rm dB})=\sigma^{2}_{(s)_{\rm dB}}\exp\left({-\frac{\|\bm{q}-\bm{q}'\|}{d_\mathrm{ref}}}\right) \,,
\end{equation}
where $\mathbb{E}(\cdot)$ denotes the expectation operator.
Lastly, when the transmitter is in $\bm{q}$, the total channel attenuation in \si{\decibel} is
\cite{bentom}
\begin{equation}\label{eq:att_db} 
({a}_{\bm{q}})_{\rm dB} = (\xi_{\bm{q}})_{\rm dB}+(s)_{\rm dB}.
\end{equation}
From \eqref{eq:att_db}, the average channel amplitude (in linear scale) can be written as 
\begin{equation} \label{eq:ch_gain}
   \tilde{g}_{\bm{q}}=10^{-\frac{({a}_{\bm{q}})_{\rm dB}}{20}}. 
\end{equation}
The signal received and sampled by Bob with sampling period $T_{\rm s}$ upon transmission of a symbol $b_k$ at time $k$ is described as \cite[Sec. 2.4.2]{tse2005fundamentals}  
\begin{equation} \label{eq:received_sig}
    r_k= \tilde{g}_{\bm{q}}h_ke^{-j\phi}b_k+w_k,
\end{equation}
where $\phi$ is the common phase term due to the signal traveled distance, $h_k \sim \mathcal{CN}(0,1)$ is a complex Gaussian circularly symmetric random variable due to the (fast) fading, and $w_k \sim \mathcal{CN}(0,\sigma_w^2)$ is the thermal noise with variance $\sigma_w^2$.

We assume that the symbol period is such that the fading coefficients $h_k$ are i.i.d. random variables. This occurs if $T_{\rm s}>T_\mathrm{c}$, where $T_\mathrm{c}$ is the coherence time, i.e., the time period over which the fading term $h_k$ in \eqref{eq:received_sig} can be considered constant. In turn, $T_\mathrm{c}$ depends on the velocity of objects in the surrounding environment: the faster the environment changes, the shorter $T_\mathrm{c}$ is. Given as the maximum speed $v_{\rm max}$ of object movement with respect to Bob and $c$ as the speed of light in the air, the Doppler frequency shift experienced by Bob is
\begin{equation}\label{ew:Doppler}
    f_\mathrm{D} =f_\mathrm{c}  \frac{v_{\rm max}}{c},
\end{equation}
and following the Clarke's model \cite{Clarke1968statistical}, the coherence time is
\begin{equation}\label{ew:coherenceTime}
    T_\mathrm{c} = \sqrt{\frac{9}{16 \pi}}\frac{1}{f_\mathrm{D}}.
\end{equation}

\subsection{Channel Gain Estimation}

As detailed in Section~\ref{sec:aut_scheme}, Bob uses the channel gain as an authentication feature. To this end, Alice transmits $K$ unit-power \ac{psk} pilot symbols $b_k$, $k=1, \ldots, K,$ with $|b_k|^2=1$. Since fading and thermal noise are two independent processes, for the transmission of independent unit power symbols, from \eqref{eq:received_sig} we have $r_k \sim \mathcal{CN}(0,\tilde{g}_{\bm{q}}^2 +\sigma_w^2)$.

Bob processes the received signal  \eqref{eq:received_sig} (which depends on the transmitter position $\bm q$) by applying the function $\mu(\cdot)$ and estimates the average channel gain as
    \begin{equation} \label{eq:filtered_sig} 
        \hat{m} = \mu(\bm{q}) =\frac{1}{K} \sum_k\left| r_k b_k^*\right|^2 = \frac{1}{K} \sum_k\left| \tilde{g}_{\bm q}h_ke^{-j\phi}+w'_k\right|^2 \,,
    \end{equation}
where $w'_k$ has the same statistics as $w_k$, since the transmitted symbols have unitary power.

\subsection{Attacker Knowledge} 
We consider the worst-case scenario for Bob, where Trudy can arbitrarily move close to Alice.  Moreover, Trudy knows:
\begin{enumerate}
    \item the Bob's location,
    \item the set $\mathcal{X}$ of possible Alice's positions,
    \item the pilot sequence $\{b_k\}$, $k=1,\ldots,K$, used by Alice to communicate with Bob.
    \item Any information acquired by Bob during the training phase (better described in Section \ref{sec:aut_scheme}).
\end{enumerate}
 
\section{Challenge-Response Physical Layer Authentication} \label{sec:aut_scheme}
\begin{figure*}
    \centering
    \subfloat[Set up: step 1) .\label{fig:relay}]{\includegraphics[width = .8\columnwidth]{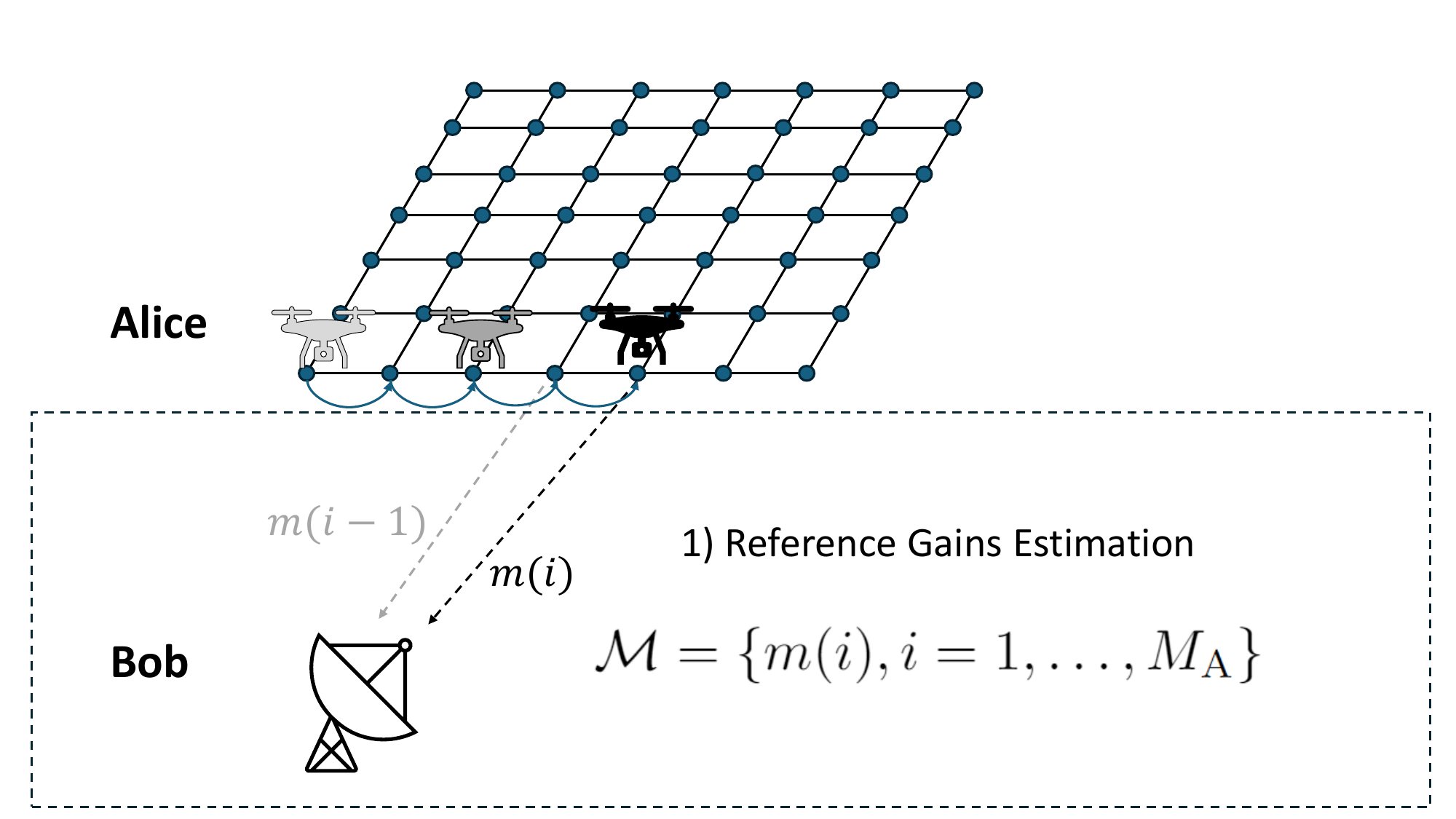} }
    \subfloat[Authentication: steps 2)-4).\label{fig:relay2}]{\includegraphics[width = .82\columnwidth]{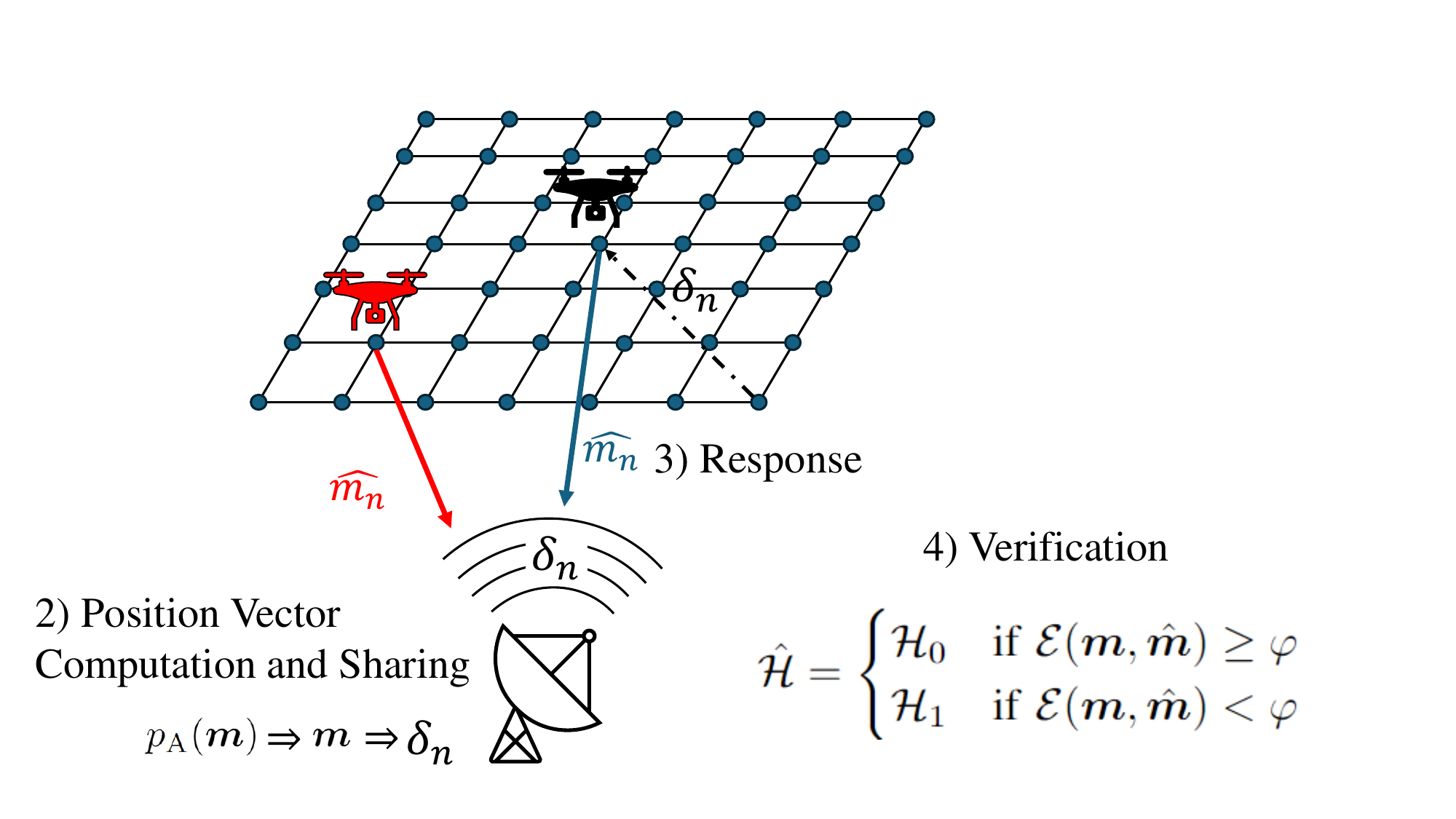} }
    \caption{Scheme of the proposed \ac{cr}-\ac{pla} protocol.}
    \label{fig:protoScheme}
\end{figure*}

We propose a physical layer authentication mechanism where Bob aims to authenticate received messages by testing the channel gains estimated from the pilot signals. Fig.~\ref{fig:protoScheme} shows the scheme of the proposed CR-PLA protocol.

The sequence of estimated channel gains is compared against the expected sequence, derived by knowing the positions from which Alice transmitted the signal. Using the channel gains to decide the authenticity of the signal allows Bob to i) avoid using the channel phase, which may lead to errors due to synchronization differences, and ii) average over noise and fading realizations to make a more reliable decision. 

We now introduce the proposed CR-\ac{pla} protocol. The protocol comprises a setup and an authentication phase. \\
\textbf{Set-up phase}:
\begin{enumerate}
    \item \emph{Reference Gains Estimation}: we assume that Alice can transmit authenticated pilot signals, for instance by using a higher-layer authentication protocol. This is a training phase, typical of the \ac{pla} schemes (e.g., \cite{Xiao2016Channel, Mazzo23Physical, Senigagliesi21Comparison, Bragagnolo21Authentication}), aimed at acquiring information used in our authentication protocol. In particular, Alice moves in all the positions in $\mathcal{X}$ and transmits authenticated pilot signals to Bob. \remembertext{k finito 1}{Then, Bob estimates the channel gain using $K=K_{\rm est}$ pilot symbols from each Alice's position using \eqref{eq:filtered_sig}. We assume this estimate to be perfect, i.e., considering $K_{\rm est}\rightarrow \infty$ in \eqref{eq:filtered_sig}. However, we will investigate in Section \ref{sec:results} the behavior of our protocol with a finite $K_{\rm est}$}.  Bob then obtains the \emph{channel gain map}
     \begin{equation} \label{eq:exp_gains_general}
    \mathcal{M} = \left\{m(i), i=1,\ldots,  M_\mathrm{A} \right\}\,.
    \end{equation}
\end{enumerate}
\textbf{Authentication phase} (to be repeated for each message transmitted by Alice):
\begin{enumerate}[resume]
    \item \emph{Position Vector Computation and Sharing}: \label{point: posComp}
     Bob randomly draws a vector $\bm{m} = [m_1, \ldots, m_N]$, $m_n \in \mathcal M$ of $N$ gains from the gain map, using the \ac{pmd} $p_{\rm A}(\bm{m})$. The design of $p_{\rm A}(\bm{m})$ is discussed in Sections~\ref{sec:game_form} and~\ref{sec:Security optimization}. Gains are mapped into corresponding positions $\bm{x}_{\rm A}=[\bm{x}_{1}, \ldots, \bm{x}_{N}]$, with $\bm{x}_n \in \mathcal{X}$, providing the average gains $\bm{m}$. Next, Bob communicates the set of positions $\bm{x}_{\rm A}$ to Alice, without disclosing them to Trudy, as better described in the following. This is considered the \emph{challenge} of the \ac{cr} protocol. 
    \item \emph{Response}: The transmission of a message by Alice is split into $N$ {\em rounds}, where at each round $n$ Alice goes to position $\bm{x}_n$ and transmits $K$ pilot symbols. Bob estimates the channel gain for each transmission and collects the estimates into vector $\hat{\bm{m}}= [ \hat{m}_1,\dots, \hat{m}_N ]$, which is considered to be the \emph{response} of our CR-PLA protocol.

    \item \emph{Verification}: Bob must decide whether the estimated vector $\hat{\bm{m}}$ matches the expected gain vector $\bm{m}$; thus, whether the received signals were transmitted by Alice or not. This is a binary decision problem. In formulas, let $\mathcal{H}_0$ and $\mathcal{H}_1$ be the legitimate (null) and the under-attack hypothesis, respectively. Let $\mathcal{H}$ the current state ($\mathcal{H}=\mathcal{H}_0$ when Alice transmits and $\mathcal{H}=\mathcal{H}_1$ when Trudy transmits). Given a test function $ \mathcal{E}(\bm{m},\hat{\bm{m}})$, the correspondent binary decision $\hat{\mathcal{H}}$ on measurement $\hat{\bm{m}}$ is
    \begin{equation}\label{eq:verification}
        \hat{\mathcal{H}} = \begin{cases}
            \mathcal{H}_0 \quad &\mbox{if }\mathcal{E}(\bm{m},\hat{\bm{m}}) \geq \varphi\,,\\
            \mathcal{H}_1 \quad &\mbox{if }\mathcal{E}(\bm{m},\hat{\bm{m}}) < \varphi\,,
        \end{cases}
    \end{equation}
    where $\varphi$ is a user-defined threshold. The choice of the test function is discussed in Section~\ref{sec:verification}.
    \end{enumerate}
 
Concerning Step~\ref{point: posComp}), several approaches can be adopted to securely communicate the position vector $\bm{x}_{\rm A}$ to Alice.
\remembertext{QuantumSecureInitPos}{As a first solution, we may assume that the initial position of Alice is secret, e.g., by having Bob randomly choose that and then secretly communicate it to Alice, for instance, using a key shared before deployment and encoding the message with a one-time pad.} Next, Bob only needs to communicate the shifts from the current to the next positions $\bm{x}_{n+1}$, i.e., $\bm{{\delta}}_n = \bm{x}_{n+1}-\bm{x}_{n}$, $n=1,\ldots, N-1$. This transmission can be public, as the shift alone does not leak information on the final position. Thus, the shift vector $\bm{\delta}$ constitutes the {challenge} of the \ac{cr} protocol to be posed to the device to be authenticated.
\remembertext{AlternativeSharing}{Note that, however, if Trudy was able to estimate accurately Alice's current location, she may also anticipate her next position by exploiting the public challenge $\bm{\delta}_n$, thus breaking the authentication protocol. A solution is then to design and secretly share the whole trajectory $\bm{x}_{\rm A}$ with Alice, either before the protocol starts or, again, using a confidential transmission (e.g., one-time pad or even wiretap coding \cite{Wyner1975wire-tap}). While this solution has a higher cost, it has relaxed assumptions on the localization capability of Trudy. Moreover, current state-of-the-art localization techniques assume the tracking device(s) to be fixed on the ground; thus, Trudy should be both fixed and equipped with additional hardware to be a threat (see  \cite{9409835,Srigrarom2021Multi-camera}).} 

Still about step 2), we assume that for each average gain $m \in \mathcal M$, there is a single corresponding position in $\mathcal X$, since the shadowing model yields a vanishing probability of having the same gain in multiple map positions.  

\subsection{Attack Model}\label{sec:att_and_def}

We consider a scenario where Trudy aims at fooling Bob by picking a sequence of $N$ positions  $\bm{y}_{\rm T}=[\bm{y}_1, \ldots, \bm{y}_N]$, where $\bm{y}_n \in \mathcal{Y} $. 

Analogously to \eqref{eq:exp_gains_general}, we define the channel gain set for the Trudy positions as  
\begin{equation} \label{eq:exp_gains_general_trudy}
    \mathcal{M}_{\rm T} = \left\{m_{\rm T}(j),j=1,\ldots, M_{\rm T}\right\}.
\end{equation}


We consider the worst case wherein the positions chosen by Trudy can be anywhere, also very close to Alice.

The Trudy attack is successful when it passes the test \eqref{eq:verification}. Hence, Trudy attack consists in i) choosing a \ac{pmd} $p_{\rm T}(\bm{m}_{\rm T})$ of a vector $\bm{m}_{\rm T}= \{m_{{\rm T},1},\ldots,m_{{\rm T},N}\}$, ii) using it to sample a vector $\bm{m}_{\rm T}$, and iii) mapping it back to a sequence of positions $\bm{y}_{\rm T}$.
In this paper, we assume the attacker can estimate the set in \eqref{eq:exp_gains_general_trudy}, e.g., by intercepting signals coming from Bob when transmitting the challenge.

\subsection{Verification Test}\label{sec:verification}

We focus on the verification problem of step 4), and we resort to decision theory to optimize the test function. In the design of the test at Bob, we assume that we do not know the attack statistics, which depend on Trudy's \ac{pmd} $p_{\rm T}(\bm{m}_{\rm T})$. Indeed, designing a test for specific statistics is, in general, sub-optimal since Trudy may then deviate from it, and choose another one that may be more effective to pass test \eqref{eq:verification}. Before delving into the definition of the test function $ \mathcal{E}(\bm{m},\hat{\bm{m}})$, let us introduce the following Lemmas.
\begin{lemma}\label{lemma:channelStat}
    When Alice transmits a large number of pilot symbols ($K \rightarrow \infty$), the estimated channel gain $\hat{m}$ with Alice in position $\bm{x}(i)$ has statistics
    \begin{equation}\label{eq:central_limit_Alice}
        \hat{m} \sim \mathcal{N}\left( m(i),\frac{m(i)^2}{K}\right), \quad \mbox{with } \; m(i) \triangleq  \tilde{g}_{\bm{x}(i)}^2+\sigma_w^2\,.
    \end{equation}
\end{lemma}
\begin{IEEEproof} 
    Proof in Appendix~\ref{app:lemma}.
\end{IEEEproof}

\begin{lemma} \label{lemma:ind_Alice}
Under the legitimate hypothesis $\mathcal{H}_0$, estimations $\hat{m}_n$ and $\hat{m}_{n'}$ taken at different times $n$ and $n'$ are statistically independent.
\end{lemma}
\begin{IEEEproof}
    Proof in Appendix~\ref{app:ind_Alice}.
\end{IEEEproof}

Now, to implement \eqref{eq:verification}, we resort to the \ac{llt} test. In particular, let us define the log-likelihood for $\mathcal{H}=\mathcal{H}_0$ given $\bm{m}$ as 
\begin{equation}\label{eq:LLTTest_long}
    \mathcal{E}'(\hat{\bm{m}},\bm m) = \log p_{\mathcal{H}_0}(\hat{\bm{m}}| \bm{m})  \,,
\end{equation}
where $p_{\mathcal{H}_0}(\hat{\bm{m}}| \bm{m})$ is the \ac{pdf} of $\hat{\bm{m}}$ given the challenge $\bm m$ when Alice is transmitting. When $\mathcal{H}=\mathcal{H}_0$, the entries of $\hat{\bm{m}}$ are fully determined by the gains $\bm{m}$ of step 2). Still, following Lemma~
\ref{lemma:ind_Alice}, estimates relative to different transmitting positions are independent. When Alice is transmitting ($\mathcal{H}=\mathcal{H}_0$), from \eqref{eq:central_limit_Alice}  we have that $\hat{m}_n = m_n + \tilde{w}_n$, where $\tilde{w}_n=\mathcal{N}\left(0,{m^2_n}/{K}\right)$. Moreover, \eqref{eq:LLTTest_long} becomes
\begin{equation}
    \begin{split}\label{eq:LLTTest}
        \mathcal{E}'(\hat{\bm{m}},\bm m) &= \log p_{\mathcal{H}_0}(\hat{\bm{m}}| \bm{m}) \\
        &= \log \prod_{n=1}^N \frac{K}{\sqrt{2\pi} m_n} \exp \left(-\frac{K(\hat{m}_n-m_n)^2}{2 m_n^2}\right)\\
        & \simeq \sum_{n=1}^N  \frac{K (\hat{m}_n-m_n)^2}{m_n^2} = \mathcal{E}(\hat{\bm{m}},\bm{m})\,,
    \end{split}
\end{equation}
where in the last approximation we neglected scalar factors that do not alter test \eqref{eq:verification}.

We remark that, differently from the likelihood ratio test, the \ac{llt} has no optimality guarantee. Still, the test is effective and is commonly used in many security applications, e.g., \cite{Chiarello21Jamming, Bragagnolo21Authentication,Peng24GLRT}. Eventually, it is also possible to implement such a test without knowing the statistical description of $\hat{\bm{m}}$ under $\mathcal{H}_1$ (see \cite{Senigagliesi21Comparison} and \cite{ardizzon2024learning}).

\section{\ac{cr}-\ac{pla} Optimization Problem} \label{sec:game_form}

The \ac{cr}-\ac{pla} mechanism introduced in the previous section has several tunable parameters, in particular the test threshold $\varphi$, and $p_{\rm A}(\bm{m})$ and $p_{\rm T} (\bm{m}_{\rm T})$ of Bob and Trudy, respectively. All these parameters have an impact on the security and energy consumed by Alice.

In terms of security, the authentication mechanism can be characterized by the \ac{fa} and \ac{md} probabilities. Let $\mathbb{P}(\cdot)$ denote the probability operator.
 \ac{fa} events occur when Alice's messages are considered fake and the corresponding probability is
    \begin{equation}
    P_{\rm fa}={\mathbb P}[\hat{\mathcal{H}} =  \mathcal{H}_1| \mathcal{H}=\mathcal{H}_0]. 
    \end{equation}
A \ac{md} event occurs when Trudy messages are considered authentic and the corresponding \ac{md} probability is
    \begin{equation}
    P_{\rm md}={\mathbb P}[\hat{\mathcal{H}} =  \mathcal{H}_0| \mathcal{H}=\mathcal{H}_1].
    \end{equation}
In terms of energy consumption by Alice, considering model \eqref{eq:energy}, the performance can be assessed through the average traveled distance over the $N$ rounds, i.e.,
\begin{equation} \label{eq: avg_distance_fin}
    \bar{D} = {\mathbb E}\left[ \sum_{n=1}^{N-1} d_{i_n,i_{n+1}}\right]\,,
\end{equation}
where $i_n$ refers to the position $\bm{x}(i_n)$, i.e., the position of Alice at round $n$ and the expectation depends on $p_{\rm A}(\bm{m})$.  

\paragraph*{Threshold Setting}
As a typical approach in hypothesis testing problems, we set a target \ac{fa} probability and the threshold $\varphi$ is chosen to meet this requirement. As 
each term of the summation in \eqref{eq:LLTTest} is the square of an independent standard Gaussian variable, $\mathcal{E}(\hat{\bm{m}},\bm{m})$ is chi-squared distributed with $N$ degrees of freedom. The \ac{fa} probability is then
\begin{equation} \label{eq:p_fa}
    P_{\rm fa} = \mathbb{P}\left(\sum_{n=1}^N  \frac{K (\hat{m}_n-m_n)^2}{m_n^2} \geq \varphi \, \Big| \, \bm{m} \right) = 1- \chi^{2}_N(\varphi)\,,
\end{equation}
where $\chi^{2}_N(\cdot)$ is the \ac{cdf} of a chi-squared variable with order $N$. Thus, we can then compute the threshold $\varphi$ of the authentication test \eqref{eq:verification} for a desired $P_{\rm fa}$ as
\begin{equation} \label{eq:chi_sq}
    \varphi =[\chi^{2}_N]^{-1} \left( 1- P_{\rm fa}\right)\,,
\end{equation}
where $[\chi^{2}_N]^{-1}(\cdot)$ is the inverse \ac{cdf} of the chi-squared distribution with order $N$.  

For the optimization of the \acp{pmd}, we consider both the \ac{md} probability and the average traveled distance as targets. However, we assume that the security performance (in terms of \ac{md} probability) is more important than the energy performance (in terms of $\bar{D}$). 

For the security performance, Bob and Trudy have conflicting targets. Trudy aims at maximizing the \ac{md} probability to obtain more effective attacks, i.e.,
\begin{equation} \label{eq_opt_trudy_new}
    p_{\rm T}^\star(\bm{m}_{\rm T})=\argmax_{p_{\rm T}(\bm{m}_{\rm T})}P_{\rm md}\,.
\end{equation}
Bob instead aims at choosing $p_{\rm A}(\bm{m})$ to minimize the \ac{md} probability. Then, for the energy performance, among all the distributions yielding the same \ac{md} probability, Bob chooses the one minimizing the average travel distance. Thus, defining the set of distributions minimizing the \ac{md} probability as
\begin{equation} \label{eq:opt_fa_sec}
    {\mathcal P}_{\rm min}=\left \{  \hat{p}_{\rm A}(\bm{m})= \argmin_{p_{\rm A}(\bm{m})}P_{\rm md} \right \}\,,
\end{equation}
we look for the \ac{pmd} minimizing the average distance in this set, i.e.,
\begin{equation} \label{eq:opt_pfa_dist}
    p^\star_{\rm A}(\bm{m})=\argmin_{p_{\rm A}(\bm{m}) \in {\mathcal P}_{\rm min} }\bar{D}\,.
\end{equation}

We observe that the optimization problems \eqref{eq_opt_trudy_new}, \eqref{eq:opt_fa_sec}, and \eqref{eq:opt_pfa_dist} are intertwined since the \ac{md} probability depends on both \acp{pmd} $p_{\rm A}(\bm{m})$ and $p_{\rm T}(\bm{m}_{\rm T})$. Now, we observe that the (average) gain estimated by Bob at round $n$ when Trudy is transmitting is
\begin{equation} \label{eq:trudy_attacco}
    \hat{m}_n = z_n =
        m_{{\rm T},n}+ \tilde{{w}}_{{\rm T},n}\,,
\end{equation}
where $m_{{\rm T},n} \in \mathcal{M}_{\rm T}$ is the expected channel gain at round $n$ and $\tilde{{w}}_{{\rm T},n} \sim \mathcal{N}\left(0,{m_{{\rm T},n}^{2}}/{K}\right) $ is the estimation error (as for Alice). Thus, combining the test function \eqref{eq:LLTTest} with \eqref{eq:trudy_attacco}, the average \ac{md} probability can be written as 
\begin{equation} \label{eq:pmd_complete}
\begin{split}
       P_{\rm md} =& \mathbb{E}_{{\bm m},{\bm m_{\rm T}}}\left[ \mathbb{P}\left(\sum_{n=1}^N  \frac{K (z_n-m_n)^2}{m_n^2} \leq \varphi|\bm{m},\bm{m}_{\rm T} \right) \right] \\
       =& \sum_{\bm m}\sum_{\bm m_{\rm T}} \mathbb{P}\left(\sum_{n=1}^N  \frac{K (z_n-m_n)^2}{m_n^2} \leq \varphi|\bm{m},\bm{m}_{\rm T} \right) \times \\ 
       &p_{\rm A}(\bm{m})p_{\rm T}(\bm{m}_{\rm T})\,, 
\end{split}
\end{equation}
which depends on both Alice's and Trudy's \acp{pmd}.

\subsection{Single and Multiple Rounds}\label{at_sec}

We now show that in the considered scenario, both Bob's and Trudy's security-optimal \ac{pmd} (i.e., problems \eqref{eq_opt_trudy_new} and \eqref{eq:opt_fa_sec}) yield independent gain realization at each round, i.e.,
    \begin{equation} \label{eq:alice_pa}
        p^\star_{\rm A}(\bm{m}) = \prod_{n=1}^N p^\star_{{\rm A},n}(m_n) \,,
    \end{equation}
    \begin{equation}\label{eq:attackFactoriz}
       p^\star_{\rm T}(\bm{m}_{\rm T}) =  \prod_{n=1}^{N} p^\star_{{\rm T},n}(m_{{\rm T},n}) \,.
    \end{equation}
In other words, both Bob and Trudy can extract independent gains across rounds without sacrificing security.

We start noting that Bob is agnostic of Trudy's attack \ac{pmd}, i.e., he does not know $p_{\rm T}(\bm{m}_{\rm T})$. Hence, there is no relation among different attacks Bob can exploit. Moreover, also following Lemma~\ref{lemma:ind_Alice}, for a given $\bm{m}$, the gain estimations in $\hat{\bm{m}}$ are independent across rounds. Thus, his choice of the position will be independent at each round, and the \ac{pmd}s constituting the set ${\mathcal P}_{\rm min}$ in \eqref{eq:opt_fa_sec} can be written as  \eqref{eq:alice_pa}. 

Now, we introduce the following theorem on Trudy's optimal strategy.
\begin{theorem}\label{th:trudyDinamic}
    For the considered scenario, given Bob's defense \ac{pmd} $p^\star_{\rm A}(\bm{m})$ as in \eqref{eq:alice_pa}, Trudy's optimal attack \ac{pmd} can be factorized as in \eqref{eq:attackFactoriz}. In other words, Trudy's best attack is to extract independent gains at each round.
\end{theorem}
\begin{IEEEproof} Proof in Appendix~\ref{app:ind_trudy}.\end{IEEEproof}

From \eqref{eq:alice_pa} and Theorem~\ref{th:trudyDinamic}, both Bob and Trudy choose their gains at each round independently. This implies that the joint optimization of the  \acp{pmd} can be done at each round. Note, however, that the \acp{pmd} can be different at each round. 
Thus, in the following, we focus on the case $N=1$, i.e., single round, whose results are applicable also for $N>1$. In this analysis, we drop the time index $n$, and so the single-round attack and defense \acp{pmd} become $p_{\rm T}(m_{\rm T})$ and $p_{\rm A}(m)$.

\section{Game Modeling and Optimization Solution}\label{sec:Security optimization}

To address the complexity of the optimization problem defined by \eqref{eq_opt_trudy_new}-\eqref{eq:opt_pfa_dist}, we first frame the security scenario as a game between Trudy and Bob. In the game, the target gains $m$ and $m_{\rm T}$ represent the {\em actions}. Trudy aims at maximizing the \ac{md} probability, while Bob aims at minimizing it. The {\em mixed strategies} of the game are described by the \acp{pmd} $p_{\rm A}(m)$ and  $p_{\rm T}(m_{\rm T})$. We then characterize the strategies that constitute a \ac{ne} of the game, that solve problems \eqref{eq_opt_trudy_new} and \eqref{eq:opt_fa_sec} jointly. Among those strategies, we then look for solutions to minimize the average distance traveled by Alice to solve \eqref{eq:opt_pfa_dist}. 

To describe the CR-PLA mechanism as a game, let us first introduce the probability vector $\bm{a}$ with length $M_{\rm A}$ and entries
    \begin{equation} \label{eq:prob_vectors_Alice}
            a_i =  p_{{\rm A}}(m(i))\,, \quad  i=1,\ldots, M_{\rm A}\,,
    \end{equation}
and probability vector $\bm{e}$ with length $M_{\rm T}$ and entries
        \begin{equation} \label{eq:prob_vectors_Trudy}
            e_j =  p_{{\rm T}}(m_{\rm T}(j))\,, \quad j=1,\ldots, M_{\rm T}\,.
    \end{equation} 

The game can then be defined as follows:
\begin{description}
    \item[Players] Bob against Trudy.
    \item[Sets of actions] set $\mathcal{M}$ and $\mathcal{M}_{\rm T}$ of the gains (and so positions) Bob and Trudy can take.
    \item[Payoff] for Trudy, the payoff obtained when Bob and Trudy play their actions (choose their positions) is the resulting \ac{md} probability $P_{\rm md}$; correspondingly, for Bob the payoff is $-P_{\rm md}$. The payoff matrix for Trudy is $\bm{P} \in [0,1]^{ M_\mathrm{A} \times M_\mathrm{T} }$, with entry (from \eqref{eq:LLTTest} and \eqref{eq:pmd_complete}) $i=1, \ldots, M_\mathrm{A}$, and $j=1, \ldots, M_\mathrm{T}$
    \begin{equation} \label{eq:pmd_matrix} 
    \begin{split}
       P_{i,j}  &=
        \mathbb{P}\left[\mathcal{N}\left(\frac{(m_{\rm T}(j)-m(i))\sqrt{K}}{m_{\rm T}(j)},1 \right)^2 \leq \varphi \frac{m(i)^2}{m_{\rm T}(j)^2}\right ]       \\ 
        &=\chi^2_{\rm NC}\left(\varphi \frac{m(i)^2}{m_{\rm T}(j)^2};\;\lambda=\frac{(m_{\rm T}(j)-m(i))^2K}{m_{\rm T}(j)^2}\right)\,,
    \end{split}
\end{equation}
where $\chi^2_{\rm NC}(\cdot;\lambda)$ is the non-central chi-squared \ac{cdf} with parameter $\lambda$. Bob's payoff matrix is instead $-\bm{P}$.
     \item[Strategies] the \acp{pmd} with which actions are taken by two players are $\bm{a}$ and $\bm{e}$, for Bob and Trudy, respectively. 
     \item[Utility function] the utility function for Trudy is defined as
    \begin{equation} \label{eq:utility_eve_mixed}
u_{\rm T}\left(\bm{a},\bm{e}\right)=\sum_{i=1}^{M_{\rm A}}\sum_{j=1}^{M_{\rm T}} P_{i,j} a_i e_j \,,
\end{equation}
i.e., her expected payoff resulting from the strategies of both players. The utility of Bob instead is $u_{\rm A}\left(\bm{a},\bm{e}\right)=-u_{\rm T}\left(\bm{a},\bm{e}\right)$. Note that the utility function of Trudy \eqref{eq:utility_eve_mixed} is equivalent to \eqref{eq:pmd_complete} (with $N=1$).
\end{description}

In game theory, a strategy is pure when the probability distribution is degenerate, e.g., for Bob $a_i=1$ for some $i$ and 0 otherwise; in all the other cases (including our game) the strategy is \emph{mixed}. Note that Bob and Trudy choose a strategy, not an action: these two concepts are the same only for a pure strategy. 

The game can be classified as a static {\em zero-sum} game of complete information. It is static because there is no time dependency, and single-shot moves fully determine the outcome of the game. It is a zero-sum game because the two players have opposite objectives, i.e., $u_{\rm A}\left(\bm{a},\bm{e}\right)=-u_{\rm T}\left(\bm{a},\bm{e}\right)$. It has complete information since the sets of actions and the payoffs are common knowledge.

\subsection{Nash Equilibrium Characterization}

  We recall that strategies $(\bm{a}^\star,\bm{e}^\star)$ are a \ac{ne} (i.e., optimal from a game theory perspective) if there is no unilateral deviation of the players leading to a better utility. In formulae, if $(\bm{a}^\star,\bm{e}^\star)$ is a \ac{ne}, then
\begin{equation}
u_{\rm A}(\bm{a}^\star,\bm{e}^\star) \geq u_{\rm A}(\bm{a},\bm{e}^\star) \quad \forall \bm{a}\, ,
\end{equation} 
and
\begin{equation} \label{eq:ne_def_1}
    u_{\rm T}(\bm{a}^\star,\bm{e}^\star)\geq u_{\rm T}(\bm{a}^\star,\bm{e}) \quad \forall \bm{e}\, .
\end{equation}

We now characterize the set of \ac{ne} strategies for the considered game. \remembertext{zero_sum_clar}{Since the game is zero-sum, let us first define two problems that we will show are strategically equivalent. Let us define the maximin problem, where Bob first optimizes his strategy and then Trudy optimizes her  strategy as} 
\begin{equation} \label{eq:maximin}
u_{\rm T, \max}^{\star}=\max_{\bm{e}}\min_{\bm{a}}u_{\rm T}(\bm{a},\bm{e}).
\end{equation} 
Similarly,  let us define the minimax problem, where Trudy first optimizes her strategy and then Bob optimizes his strategy as  
\begin{equation} \label{eq:minimax}
u_{\rm T, \min}^{\star}=\min_{\bm{a}}\max_{\bm{e}}u_{\rm T}(\bm{a},\bm{e}).
\end{equation} 
Let us also define Trudy's strategies $\bm{e}_{\rm max}$ and $\bm{e}_{\rm min}$ that lead to the $u_{\rm T, \max}^{\star}$ \eqref{eq:maximin} and $u_{\rm T, \min}^{\star}$ \eqref{eq:minimax}, respectively. 

Since the game is zero-sum, the following results hold:
\begin{enumerate}
    \item As the set of actions is finite, at least one \ac{ne} exists in mixed strategies \cite[Sec. 6.4]{tadelis2013game}.    
    \item From the minimax theorem \cite{v1928theorie}, we have
    \begin{equation} \label{eq:minimax_eq}
        \bm{e}^\star = \bm{e}_{\rm max} = \bm{e}_{\rm min}, \quad u^{\star} = u_{\rm T, \min}^{\star} = u_{\rm T, \max}^{\star}\,,
    \end{equation}
    \remembertext{zero_sum_clar_2}{thus \eqref{eq:maximin} and \eqref{eq:minimax} are equivalent.} 
    \item 
    \begin{lemma} \label{lemma:same_payoff}
All \acp{ne} yield the same utility.
\end{lemma}
\begin{IEEEproof}
    Proof in Appendix~\ref{app:same_payoff}.
\end{IEEEproof}
\end{enumerate}

\subsection{Game-Based Optimization Problem} \label{sec:strategy_opt}
From \eqref{eq:minimax_eq}, we can compute the utility of any \ac{ne} by solving the maximin problem \eqref{eq:maximin} via linear programming as shown in \cite[Sec. 3.10]{washburn2014two}. In particular, by letting $\bm{P}_i^T$ to be the $i$-th row of $\bm{P}$, the \ac{ne} utility is the solution of the following optimization problem
\begin{equation} \label{eq:opt_problem}
\begin{aligned}
u^{\star}=   \max_{u} \quad & u\\
\textrm{s.t.} 
\quad & u \leq  \bm{P}_i^T \bm{e}, \qquad \forall i \in 1, \ldots, M_{\rm A},\\
& \sum_{j=1}^{M_{\rm T}} e_j = 1,  \quad e_j \geq 0, \forall j.
\end{aligned}
\end{equation}
Note that the utility of the game $u^{\star}$ coincides with the average MD probability of \eqref{eq:pmd_complete} and $\bm{e}^{\star}$ (i.e., the solution of \eqref{eq:opt_problem} achieving $u^{\star}$) is an optimal solution for Trudy. 

As Bob's utility is the opposite of Trudy, we now have a characterization of the set ${\mathcal P}_{\rm min}$ in \eqref{eq:opt_fa_sec}, as the set of strategies solving \eqref{eq:opt_problem} with $-\bm{P}^T$ in place of $\bm{P}$ (see \cite[Sec. 3.10]{washburn2014two}), thus we have
\begin{equation} \label{eq:opt_fa_sec_gt}
    {\mathcal P}_{\rm min}=\{ \bm{a}: -\bm{P}^T \bm{a} \geq u^\star \mathds{1} \}\,,
\end{equation}
where $\mathds{1}$ is the vector with all entries equal to 1.
\remembertext{complexity}{We remark that, thanks to Theorem \ref{th:trudyDinamic}, the \ac{ne} for the zero-sum game with the utility is the \ac{md} over multiple rounds can be derived straight from the \ac{ne} obtained in~\eqref{eq:opt_fa_sec_gt}.}
We will use this result to find the minimum-distance best response for Bob in the next subsections.

\remembertext{dist among ne}{Note that distance minimization has no effect on Bob and Trudy's strategy. In fact, all the approaches presented in the next subsections provide solutions within the \ac{ne} set of \eqref{eq:opt_fa_sec_gt}. Consequently, distance minimization finds the energy optimal \ac{ne} among all \acp{ne}. According to Lemma~\ref{lemma:same_payoff}, since this is a zero-sum game, all \acp{ne} yield the same payoff; thus, Trudy cannot use this information to design a better strategy. Finally, we note that we could have included both the safety (\ac{md} probability minimization) and the energy objectives in the payoff function for Bob. However, this would have caused two major problems: a) mixing the two goals (security and energy reduction) is problematic because they are associated with incomparable metrics (misdetection probability and energy); b) the resulting game would not be zero-sum, making the analysis of \acp{ne} much more complex. Therefore, we decided to prioritize the security objective by considering only \acp{ne} that minimize Bob's \ac{md} probability, which yields a zero-sum game, and then selecting the \ac{ne} that minimizes energy.}

\subsection{Multi-Round Multiple-NE Solution}\label{sec:MNE}

We now consider again the optimization problem \eqref{eq_opt_trudy_new}-\eqref{eq:opt_pfa_dist} with multiple rounds.  \remembertext{eq_complicata}{We denote as $\bm{a}(i,n)$ the \ac{pmd} to use when Alice is located in position $\bm{x}(i)$ at round $n$. Specifically, each element $a_{j}(i,n)$ represents the probability of Alice moving to position $\bm{x}(j)$ at round $(n+1)$ given that she is in position $\bm{x}(i)$ at round $n$. We denote as $\bm{A}(n)$ the $M_{\rm A} \times M_{\rm A}$ matrix whose column $i$ is the vector $\bm{a}(i,n)$, i.e., $\bm{A}(n)_{\cdot,i} = \bm{a}(i,n)$. Next, we collect in $\bm{A}= [ \bm{A}(1),\ldots, \bm{A}(N)]$ the optimal strategies for each position, at each round. The probability of obtaining the sequence of position (indexes) $\mathcal{I}=\{i_1,\ldots, i_N\}$ is $ p_{\mathcal I}(\mathcal{I},\bm{A})=\prod_{n=1}^{N-1} a_{i_{n+1}}(i_n,n)$. 

Now, since positions and gains have a one-to-one mapping (see \eqref{eq:exp_gains_general}), for each sequence of positions $\mathcal{I}$ we have a corresponding vector $\bm m$ and $p_{\mathcal I}(\mathcal{I},\bm{A}) = p_{\rm A}(\bm{m})$. 

Moreover, the total distance associated with the sequence $\mathcal{I}$ is $\sum_{n=1}^{N-1} d_{i_n,i_{n+1}}$ where $ d_{i_n,i_{n+1}}$ is defined in \eqref{eq:def_dist}. Thus, the average covered distance in \eqref{eq:opt_pfa_dist} can be written as  $\bar{D} = \sum_{\mathcal{I}=\{i_1,\ldots,i_N\}}  \left[\sum_{n=1}^{N-1} d_{i_n,i_{n+1}} \right] p_{\mathcal I}(\mathcal{I},\bm{A})$. About the constraints, we simply force all the probabilities to be Nash Equilibrium.

Given these considerations, the optimization problem to find the optimal multi-round strategy $\bm{A}^\star$ that corresponds to $p^\star_{\rm A}(\bm{m})$ in \eqref{eq:opt_pfa_dist} can be rewritten as}  
    \begin{equation} \label{eq:opt_NE}
        \begin{aligned}
        \bm{A}^\star=
        {\rm arg} \min_{\bm{A}} \;  \hspace{-2mm} &\sum_{\mathcal{I}=\{i_1,\ldots,i_N\}}  \left[\sum_{n=1}^{N-1} d_{i_n,i_{n+1}} \right] p_{\mathcal I}(\mathcal{I},\bm{A}) \\      
        \textrm{s.t.} \;\;& \forall i \in 1, \ldots, M_{\rm A} \,, \forall n:  \\
       \quad & u^{\star}\mathds{1} \leq  -\bm{P}^T \bm{a}(i,n) \hspace{0.5cm} \\
        & \sum_{i'=1}^{M_{\rm A}} a_{i'}(i,n)= 1, \quad \bm{a}(i,n) \geq \bm{0}. \\
        \end{aligned}
    \end{equation}
We denote the solution of problem \eqref{eq:opt_NE} as \ac{mrm}-\ac{ne} solution. \remembertext{commento_1.5}{Since the objective function includes $p_{\mathcal I}(\mathcal{I},\bm{A}) $, which is obtained as the product of the optimization variables, we have that problem \eqref{eq:opt_NE} is non-convex.}

\remembertext{complexityDistance}{
Concerning the computational complexity of \eqref{eq:opt_NE}, we note the problem is non-convex. We also note that problem \eqref{eq:opt_NE} has a large number of variables ($M_{\rm A}^2N$) and the number of operations required to evaluate the objective function is $(2N+1)M_{\rm A}^N$, i.e. an exponential number of operations. Finally, the structure of the problem does not suggest any algorithm to solve it efficiently. 
Thus, in the following, we present a relaxed version of \eqref{eq:opt_NE} that can be solved with lower complexity.
}

\subsection{Single-Hop Multiple-NE Solution}
As we will see in Section~\ref{sec:results}, even with a small number of rounds ($N=2,3$), we achieve a low \ac{md} probability. This suggests that, instead of optimizing over multiple rounds, we could neglect the effect of choosing a position at round $n$ on the distance traveled on subsequent rounds. In other words, we could neglect time in the optimization and apply the same strategy at each round (see \cite{ardizzon2024energy}). Thus, we simplify \eqref{eq:opt_NE} by finding a \ac{ne} that depends solely on Alice's current position, thus, we have $\bm{a}^\star(i,n)=\bm{a}^\star(i)$.  For each position $\bm x (i)$, the corresponding NE that minimizes the average distance traveled when the drone is in that position is 
    \begin{equation} \label{eq:greedy_solution}
        \begin{aligned}
        \bm{a}^\star(i)= {\rm arg}\min_{\bm{a}(i)} \quad &\sum_{i'=1}^{M_{\rm A}} d_{i,i'} a_i'(i)\\
        \textrm{s.t.} \;\; &    u^{\star}\mathds{1} \leq  -\bm{P}^T \bm{a}(i) \hspace{0.5cm} \\
        & \sum_{i'=1}^{M_{\rm A}} a_{i'}(i)= 1 \,, \bm{a}(i) \geq \bm{0}\,.
        \end{aligned}
    \end{equation}
We denote this problem as \ac{shm}-\ac{ne} solution. 

Note that to find the optimal strategy $\bm{A}^\star(n)=\bm{A}^\star$ we need to solve \eqref{eq:greedy_solution} $ M_{\rm A}$ times (one for each starting position $\bm x (i)$), but each solution is optimal and much easier to compute than \eqref{eq:opt_NE}, since \eqref{eq:greedy_solution} is a simple and fast linear program.

\begin{table}
\centering
   \caption{Simulation Parameters}
\label{tab:SimParam}

    \begin{tabular}{ |>{\centering\arraybackslash}p{1cm}||p{3.5cm}|>{\centering\arraybackslash}p{1.2cm}|  }
         \hline
         Symbol & Description & Default value\\ 
         \hline
         $K$ & Number of pilots symbols  &   50\\
         $N$ & Number of rounds  & 1 \\
         $\sigma_{(s)_{\rm dB}}$&   Shadowing STD  & 10 \si{\decibel}\\
         $h$&   Altitude  & \SI{50}{\meter} \\
         $M_{\rm A}$&   Number of Alice positions  & 15 $\times$ 15 \\
         $M_{\rm T}$&   Number of Trudy positions  & 15 $\times$ 15 \\
         $f_{\rm c}$ & Carrier frequency & \SI{2.4}{\giga\hertz} \\
         \hline
\end{tabular}
\end{table}

\subsection{Single \ac{ne}  Solution}\label{sec:SNE}

In this last approach, we find a single \ac{ne} $\bm{a}^\star$ that minimizes the average traveled distance, without taking into account the current and next positions, thus in a greedy fashion. Consequently we have $\bm{a}^\star(i,n)=\bm{a}^\star$. Let us define the distance matrix $\bm{D} \in \mathbb{R}^{ M_\mathrm{A} \times  M_\mathrm{A}}$, whose entries (from \eqref{eq:def_dist}) are 
\begin{equation} \label{eq:dist_matrix_definition}
    \bm{D}_{i,i'} = d_{i,i'} \quad \mbox{for } i,i'=1, \ldots , M_\mathrm{A}\,.
\end{equation}

Given a mixed strategy $\bm{a}$ for Bob, we compute the average traveled distance as
\begin{equation} \label{eq.avg_distance}
    \begin{split}
        \bar{D}(\bm{D},\bm{a})&= \sum_{i=1}^{M_{\rm A}} \sum_{i'=1}^{M_{\rm A}} d_{i,i'} a_ia_{i'} \\
         &=\bm{a}^T \bm{D} \bm{a}\,.
    \end{split}
\end{equation}
The optimization problem thus becomes
\begin{equation} \label{eq:opt_problem_distance}
\begin{aligned}
\bm{a}^{\star}=\arg \min_{\bm{a}} \quad &   \bm{a}^T \bm{D} \bm{a}\\
\textrm{s.t.} 
       \quad & u^{\star} \mathds{1} \leq  -\bm{P}^T \bm{a} \hspace{0.5cm} \\
        & \sum_{i'=1}^{M_{\rm A}} a_{i'}= 1 \,, \bm{a} \geq \bm{0}\,.
\end{aligned}
\end{equation}
We denote problem \eqref{eq:opt_problem_distance} as \ac{sne} solution.  

\remembertext{complexity single ne new}{
Since $\bm{D}$ is not positive semidefinite, \eqref{eq:opt_problem_distance} is not convex. Still, the structure of \eqref{eq:opt_problem_distance} is much simpler than \eqref{eq:opt_NE} as it suggests the usage of sequential least squares programming (SLSQP) \cite{boggs1995sequential} that exploits the knowledge of the Hessian matrix, which in \eqref{eq:opt_problem_distance} is clearly $\bm D$.}

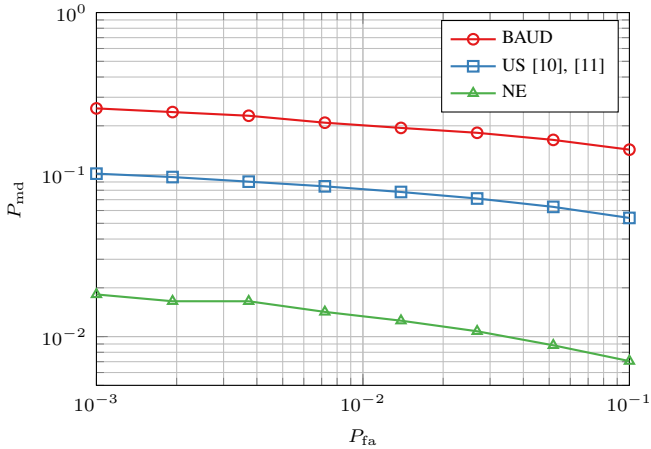
\begin{figure}
    \centering
%
%
\definecolor{mycolor1}{RGB}{228,26,28}%
\definecolor{mycolor2}{RGB}{55,126,184}%
\definecolor{mycolor3}{RGB}{77,175,74}%
\begin{tikzpicture}

\begin{axis}[%
width=\fwidth,
height=0.7\fheight,
scale only axis,
xmode=log,
xmin=0.001,
xmax=0.1,
xminorticks=true,
xlabel={$P_{\rm fa}$},
ymode=log,
ymin=0.005,
ymax=1,
yminorticks=true,
ylabel={$P_{\rm md}$},
xmajorgrids,
xminorgrids,
ymajorgrids,
yminorgrids,
legend style={legend cell align=left, align=left}
]
\addplot [color=mycolor1, thick,mark=o, mark options={solid, mycolor1}]
  table[row sep=crcr]{%
0.001	0.25610025\\
0.00193069772888325	0.24304975\\
0.00372759372031494	0.2306995\\
0.00719685673001152	0.209028\\
0.0138949549437314	0.19409425\\
0.0268269579527972	0.1810745\\
0.0517947467923121	0.1636815\\
0.1	0.142501\\
};
\addlegendentry{BAUD}

\addplot [color=mycolor2, thick,mark=square, mark options={solid, mycolor2}]
  table[row sep=crcr]{%
0.001	0.101302000713858\\
0.00193069772888325	0.0965304043859491\\
0.00372759372031494	0.0904021054525384\\
0.00719685673001152	0.0845967706875838\\
0.0138949549437314	0.0780722686604396\\
0.0268269579527972	0.0711588463452284\\
0.0517947467923121	0.0631820332417289\\
0.1	0.0540019231725464\\
};
\addlegendentry{US \cite{Mazzo23Physical},\cite{ardizzon2024energy}}

\addplot [color=mycolor3, thick, mark=triangle, mark options={solid, mycolor3}]
  table[row sep=crcr]{%
0.001	0.018203\\
0.00193069772888325	0.016526\\
0.00372759372031494	0.0165215\\
0.00719685673001152	0.01420375\\
0.0138949549437314	0.012536\\
0.0268269579527972	0.0107845\\
0.0517947467923121	0.00883425\\
0.1	0.00705875\\
};
\addlegendentry{NE}

\end{axis}
\end{tikzpicture}%
    \caption{\ac{det} curve of the proposed protocol with different strategies and $N=2$ rounds.}
    \label{fig:non_opt}
\end{figure}

\section{Numerical Results}\label{sec:results}
 
In this section, we report simulation results of the proposed \ac{cr}-\ac{pla} with optimized strategies, and compare them with existing techniques. \remembertext{grid_desc}{We consider a scenario wherein both Trudy and Alice move on an horizontal square grid of dimension $\SI{420}{\meter} \times \SI{420}{\meter}$ at a fixed height $h$, with 15 possible values of the position on each ($x$ and $y$) axis.} \remembertext{new variance}{The values of shadowing STDs $\sigma_{(s)_{\rm dB}}$ taken from~\cite{sharma2018study} refer to a ray-tracing simulation used to model a channel at \SI{2.4}{\giga\hertz} in different environments. In particular, values $\sigma_{(s)_{\rm dB}} \simeq 10,13$, and $\SI{16}{\decibel}$ refer to urban and suburban \ac{nlos} scenarios, respectively. On the other hand, $\sigma_{(s)_{\rm dB}} \simeq\SI{6}{\decibel}$ is taken from \cite{chandrasekharan2015propagation} where the authors conducted field
measurements in rural areas at \SI{2.4}{\giga\hertz}.} Table~\ref{tab:SimParam} summarizes the main simulation parameters; if not otherwise stated, the parameters used in the simulations are reported in the column {\em Default value}. 

\begin{figure} 
    \centering
%
%
\definecolor{mycolor1}{RGB}{228,26,28}%
\definecolor{mycolor2}{RGB}{55,126,184}%
\definecolor{mycolor3}{RGB}{77,175,74}%
\definecolor{mycolor4}{RGB}{255,127,80}%
\begin{tikzpicture}[ every plot/.style={thick}]

\begin{axis}[%
width=\fwidth,
height=0.7\fheight,
at={(0.758in,0.481in)},
scale only axis,
xmode=log,
xmin=0.0001,
xmax=1,
xminorticks=true,
xlabel={$P_{\rm fa}$},
ymode=log,
ymin=1e-05,
ymax=1,
yminorticks=true,
ylabel={$P_{\rm md}$},
xmajorgrids,
xminorgrids,
ymajorgrids,
yminorgrids,
legend style={legend cell align=left, align=left,at={(0.03,0.5)},anchor=west}
]

\addplot [color=mycolor4,thick, mark=star, mark options={solid, mycolor4}]
  table[row sep=crcr]{%
0.0001	0.208147290318366\\
0.000562341325190349	0.180956851576564\\
0.00316227766016838	0.152036940529626\\
0.0177827941003892	0.123992168761925\\
0.1	0.0875214322140757\\
0.2825	0.0583402577377432\\
0.455	0.0409521483375116\\
0.6275	0.0267461122741508\\
0.8	0.0140039899378374\\
0.9	0.00695193263671459\\
0.92475	0.00522598393778311\\
0.9495	0.00350451595585971\\
0.97425	0.00178608560098524\\
0.999	6.91824124543247e-05\\
};
\addlegendentry{$\sigma_{(s)_{\rm dB}}=6$ \si{\decibel}}

\addplot [color=mycolor1,thick, mark=square, mark options={solid, mycolor1}]
  table[row sep=crcr]{%
0.0001	0.17690812442379\\
0.000562341325190349	0.152552737252718\\
0.00316227766016838	0.130612499354946\\
0.0177827941003892	0.106006398106752\\
0.1	0.074820808144174\\
0.2825	0.0497928100070259\\
0.455	0.0349094719333417\\
0.6275	0.0227772396593468\\
0.8	0.0119201513645131\\
0.9	0.00591633388468498\\
0.92475	0.00444734873723599\\
0.9495	0.00298230980333395\\
0.97425	0.0015199243392757\\
0.999	5.88711445456635e-05\\
};
\addlegendentry{$\sigma_{(s)_{\rm dB}}=10$ \si{\decibel}}

\addplot [color=mycolor2,thick, mark=triangle, mark options={solid, mycolor2}]
  table[row sep=crcr]{%
0.0001	0.133962725311744\\
0.000562341325190349	0.11632760944939\\
0.00316227766016838	0.098480568561753\\
0.0177827941003892	0.0794082662974354\\
0.1	0.0568520408705434\\
0.2825	0.0382868048132908\\
0.455	0.0269925564526704\\
0.6275	0.0176743813882766\\
0.8	0.00926804674759431\\
0.9	0.00460267165122591\\
0.92475	0.00346014269643612\\
0.9495	0.00232045126711375\\
0.97425	0.00118265005041708\\
0.999	4.58123449551616e-05\\
};
\addlegendentry{$\sigma_{(s)_{\rm dB}}=13$ \si{\decibel}}

\addplot [color=mycolor3,thick, mark=o, mark options={solid, mycolor3}]
  table[row sep=crcr]{%
0.0001	0.116489135216096\\
0.000562341325190349	0.103035038656635\\
0.00316227766016838	0.0872482712763825\\
0.0177827941003892	0.071578899893607\\
0.1	0.0510339113072291\\
0.2825	0.0344102287866878\\
0.455	0.0242771060780296\\
0.6275	0.0159054142399764\\
0.8	0.00834383512701747\\
0.9	0.00414417832627431\\
0.92475	0.00311552048429459\\
0.9495	0.00208936273265817\\
0.97425	0.00106488674304904\\
0.999	4.12482120844551e-05\\
};
\addlegendentry{$\sigma_{(s)_{\rm dB}}=16$ \si{\decibel}}

\addplot [color=mycolor3,thick, only marks,mark=*, mark options={solid, black,scale=0.4}]
  table[row sep=crcr]{%
0.0001	0.177494\\
0.000562341325190349	0.152102\\
0.00316227766016838	0.129976\\
0.0177827941003892	0.104964\\
0.1	0.074928\\
0.2825	0.049656\\
0.455	0.034834\\
0.6275	0.02288\\
0.8	0.011852\\
0.9	0.005896\\
0.92475	0.00451\\
0.9495	0.003026\\
0.97425	0.00155\\
0.999	5.4e-05\\
};
\addlegendentry{Simulated}

\addplot [color=mycolor3, only marks, mark=*, mark options={solid, black,scale=0.4}, forget plot]
  table[row sep=crcr]{%
0.0001	0.115204\\
0.000562341325190349	0.102598\\
0.00316227766016838	0.087276\\
0.0177827941003892	0.072398\\
0.1	0.050526\\
0.2825	0.034568\\
0.455	0.024682\\
0.6275	0.015994\\
0.8	0.008376\\
0.9	0.004234\\
0.92475	0.003046\\
0.9495	0.002198\\
0.97425	0.001146\\
0.999	4.8e-05\\
};
\addplot [color=mycolor3, only marks, mark=*, mark options={solid, black,scale=0.4}, forget plot]
  table[row sep=crcr]{%
0.0001	0.208166\\
0.000562341325190349	0.180956\\
0.00316227766016838	0.153102\\
0.0177827941003892	0.124564\\
0.1	0.087474\\
0.2825	0.05851\\
0.455	0.04052\\
0.6275	0.02708\\
0.8	0.013964\\
0.9	0.006984\\
0.92475	0.004964\\
0.9495	0.003478\\
0.97425	0.001772\\
0.999	7e-05\\
};
\end{axis}
\end{tikzpicture}%
    \caption{\remembertext{new pmd}{\Ac{det} curves for $\sigma_{(s)_{\rm dB}}=6$, $10$, $13$, and $\SI{16}{\decibel}$ for simulated and theoretical analysis.}}
    \label{fig: pmd_pfa_complete}
\end{figure}

\subsection{Comparison with Other Approaches} \label{par:comparison}
We first compare the security performance of our scheme with other solutions present in the literature. In particular, we consider the uniform strategy (US) proposed in \cite{Mazzo23Physical} and \cite{ardizzon2024energy}, where Bob and Trudy uniformly pick a gain over the set of possible values of the map. We also consider the case in which Bob still picks the gain uniformly at random, while Trudy uses the strategy that maximizes the \ac{md} probability: this is the best attack against the uniform defense strategy proposed in \cite{Mazzo23Physical}, \cite{ardizzon2024energy}) and we denote it as the best attack against the uniform defense (BAUD) strategy.

Note that the various solutions discussed in Section~V achieve all the same security performance since we always consider \acp{ne} of the game, and all \acp{ne} yield the same \ac{md} probability.

Fig.~\ref{fig:non_opt} reports the \acf{det} curves,  i.e., \ac{md} probability as a function of the \ac{fa} probability,  considering the proposed game theoretic approach (denoted with NE in the plot) against the US and BAUD strategies. The number of rounds is $N=2$, while the other parameters are reported in Table~I. We confirm that the \ac{ne} provides the lowest \ac{md} probability, as the uniform strategy is not the best for Bob. Moreover, we see that the uniform strategy for Bob is particularly vulnerable to the best attack by Trudy. We also observe that we are able to reduce the \ac{md} probability by one order of magnitude with just two rounds.

\begin{figure}
    \centering
%
%
\definecolor{mycolor1}{RGB}{228,26,28}%
\definecolor{mycolor2}{RGB}{55,126,184}%

\begin{tikzpicture}[ every plot/.style={thick}]

\begin{axis}[%
width=\fwidth,
height=0.7\fheight,
at={(0.758in,0.481in)},
scale only axis,
xmode=log,
xmin=100,
xmax=10000,
xminorticks=true,
xlabel={$K_{\rm est}$},
ymode=log,
yminorticks=true,
ylabel={$P_{\rm md}$},
xlabel style={font=\scriptsize},ylabel style={font=\scriptsize},
xmajorgrids,
xminorgrids,
ymajorgrids,
yminorgrids,
legend style={legend cell align=left, align=center, draw=white!15!black, legend columns = 2}
]
\addplot [color=mycolor1,dashed, mark=o, mark options={solid, mycolor1}, forget plot]
  table[row sep=crcr]{%
100	0.164758509938162\\
193.069772888325	0.164758509938162\\
372.759372031494	0.164758509938162\\
719.685673001152	0.164758509938162\\
1389.49549437314	0.164758509938162\\
2682.69579527972	0.164758509938162\\
5179.47467923121	0.164758509938162\\
10000	0.164758509938162\\
};

\addplot [color=mycolor2, mark=o, mark options={solid, mycolor2}, forget plot]
  table[row sep=crcr]{%
100	0.107987233983303\\
193.069772888325	0.107987233983303\\
372.759372031494	0.107987233983303\\
719.685673001152	0.107987233983303\\
1389.49549437314	0.107987233983303\\
2682.69579527972	0.107987233983303\\
5179.47467923121	0.107987233983303\\
10000	0.107987233983303\\
};

\addplot [color=mycolor1, dashed, mark = square, mark options={solid, mycolor1}, forget plot]
  table[row sep=crcr]{%
100	0.24317232833439\\
193.069772888325	0.200665449266305\\
372.759372031494	0.184784107116105\\
719.685673001152	0.17481928121365\\
1389.49549437314	0.169701259501928\\
2682.69579527972	0.167826262773009\\
5179.47467923121	0.16634335986283\\
10000	0.166311123140707\\
};

\addplot [color=mycolor2, mark = square, mark options={solid, mycolor2}, forget plot]
  table[row sep=crcr]{%
100	0.123164080312963\\
193.069772888325	0.116404298699716\\
372.759372031494	0.112575181151168\\
719.685673001152	0.110451947509696\\
1389.49549437314	0.10936099435198\\
2682.69579527972	0.10871378089801\\
5179.47467923121	0.108395126983199\\
10000	0.108221788721062\\
};
\addplot [color=mycolor1, dashed, mark=triangle, mark options={solid, mycolor1}, forget plot]
  table[row sep=crcr]{%
100	0.315799616566479\\
193.069772888325	0.256926977752611\\
372.759372031494	0.213980944276558\\
719.685673001152	0.188561456085437\\
1389.49549437314	0.176243092573892\\
2682.69579527972	0.172070984795881\\
5179.47467923121	0.168371843818771\\
10000	0.168298468162046\\
};
\addplot [color=mycolor2, mark=triangle, mark options={solid, mycolor2}, forget plot]
  table[row sep=crcr]{%
100	0.143933113642908\\
193.069772888325	0.128238156839203\\
372.759372031494	0.119256844794414\\
719.685673001152	0.11392404711308\\
1389.49549437314	0.11140289758877\\
2682.69579527972	0.10995739075714\\
5179.47467923121	0.109120898993411\\
10000	0.10864496572815\\
};

\addlegendimage{color=mycolor1, dashed}
\addlegendentry{$P_{\rm fa}= 10^{-3}$}

\addlegendimage{color=mycolor2}
\addlegendentry{$P_{\rm fa}= 10^{-1}$}

\addlegendimage{color=gray, mark = o}
\addlegendentry{$K_{\rm est} = \infty$}

\addlegendimage{color=gray, mark = square}
\addlegendentry{Sym}

\addlegendimage{color=gray, mark = triangle}
\addlegendentry{Asym}

\end{axis}

\end{tikzpicture}%
\caption{ \remembertext{sensitivity plot}{\Ac{det} curves for different number of pilot symbols in the set-up phase, $K_{\rm est}$. Circles for perfect estimation $K_{\rm est} = \infty$, triangles refer to the Asym scenario, and squares refer to the Sym scenario.}}
    \label{fig: non_opt_stim}
\end{figure}

\subsection{Impact of \ac{cr}-\ac{pla} Parameters} \label{subsec: pla_param}

We now investigate the effects of several parameters of the proposed technique on its performance. 

\paragraph*{Shadowing variance $\sigma_{(s)_{\rm dB}}$}
First, Fig.~\ref{fig: pmd_pfa_complete} compares the security performance in environments characterized by different shadowing variances. The figure shows the \ac{det} curves, for $N=1$, $\sigma_{(s)_{\rm dB}}=6$, $10$, $13$, and $\SI{16}{\decibel}$. Lines are obtained with the closed-form formulas of the probability, while black markers show the results obtained with Monte Carlo simulations. We note a perfect agreement between the analytical and numerical results. \remembertext{fig3comm}{When comparing the different variances of the shadowing, we note that the \ac{md} probability decreases with $\sigma_{(s)_{\rm dB}}$ (for a fixed \ac{fa} probability). This is because a high $\sigma_{(s)_{\rm dB}}$ value increases the map diversity, i.e., positions at the same distance from the receiver have different gains, thus it is harder for Trudy to guess a position leading to the same gain as Alice to fool Bob.} 

In the next figures, we will keep the shadowing variance fixed to $\sigma_{(s)_{\rm dB}} =\SI{10}{\decibel}$.

\begin{figure}
    \centering
%
%
\definecolor{mycolor1}{RGB}{228,26,28}%
\definecolor{mycolor2}{RGB}{55,126,184}%
\definecolor{mycolor3}{RGB}{77,175,74}%
\begin{tikzpicture}[ every plot/.style={thick}]

\begin{axis}[%
width=\fwidth,
height=0.7\fheight,
at={(0.758in,0.481in)},
scale only axis,
xmode=log,
xmin=0.0001,
xmax=0.1,
xminorticks=true,
xlabel={$P_{\rm fa}$},
ymode=log,
ymin=0.0516401618030727,
ymax=0.164866402166415,
yminorticks=true,
ylabel={$P_{\rm md}$},
xmajorgrids,
xminorgrids,
ymajorgrids,
yminorgrids,
legend style={legend cell align=left, align=left}
]
\addplot [color=mycolor1,thick, mark=o, mark options={solid, mycolor1}]
  table[row sep=crcr]{%
0.0001	0.164866402166415\\
0.000143844988828766	0.158820248513939\\
0.000206913808111479	0.154490621582363\\
0.000297635144163132	0.151368536582481\\
0.00042813323987194	0.147947820226305\\
0.000615848211066027	0.144027331597675\\
0.000885866790410082	0.138421778851458\\
0.00127427498570313	0.133456323506608\\
0.00183298071083244	0.129414025608045\\
0.00263665089873036	0.12546709528848\\
0.00379269019073225	0.120160312178111\\
0.00545559478116851	0.115255741106237\\
0.00784759970351461	0.110557488658343\\
0.0112883789168469	0.105389175825016\\
0.0162377673918872	0.100137888738748\\
0.0233572146909012	0.0947777691128727\\
0.0335981828628378	0.0889822489449115\\
0.0483293023857175	0.0832674621790851\\
0.0695192796177561	0.0768848463735295\\
0.1	0.0700737255092294\\
};
\addlegendentry{$K= 50$}

\addplot [color=mycolor2,thick, mark=square, mark options={solid, mycolor2}]
  table[row sep=crcr]{%
0.0001	0.136518674008491\\
0.000143844988828766	0.133218579198593\\
0.000206913808111479	0.1304609438459\\
0.000297635144163132	0.127364316161466\\
0.00042813323987194	0.124494554011684\\
0.000615848211066027	0.120357684474968\\
0.000885866790410082	0.116833078804893\\
0.00127427498570313	0.113138342264374\\
0.00183298071083244	0.109888376077758\\
0.00263665089873036	0.105946687213459\\
0.00379269019073225	0.102259868658756\\
0.00545559478116851	0.0981983583982899\\
0.00784759970351461	0.0943695170114103\\
0.0112883789168469	0.0899625741561813\\
0.0162377673918872	0.0858641508457863\\
0.0233572146909012	0.0811512698770141\\
0.0335981828628378	0.0766289000179883\\
0.0483293023857175	0.0714934105827023\\
0.0695192796177561	0.0661443358550451\\
0.1	0.0603736442379412\\
};
\addlegendentry{$K= 70$}

\addplot [color=mycolor3,thick, mark=triangle, mark options={solid, mycolor3}]
  table[row sep=crcr]{%
0.0001	0.114500684967234\\
0.000143844988828766	0.111833474820242\\
0.000206913808111479	0.109686969778777\\
0.000297635144163132	0.106821012504839\\
0.00042813323987194	0.104342516184163\\
0.000615848211066027	0.101305233643181\\
0.000885866790410082	0.09842455259585\\
0.00127427498570313	0.0956910479628794\\
0.00183298071083244	0.0929639893786742\\
0.00263665089873036	0.0895029459348843\\
0.00379269019073225	0.0866074484394269\\
0.00545559478116851	0.0835567172067854\\
0.00784759970351461	0.0799248658181179\\
0.0112883789168469	0.0767498547586852\\
0.0162377673918872	0.0729519345435328\\
0.0233572146909012	0.069323280426444\\
0.0335981828628378	0.0652947485274341\\
0.0483293023857175	0.0610535137286144\\
0.0695192796177561	0.056512908159935\\
0.1	0.0516401618030727\\
};
\addlegendentry{$K= 100$}

\end{axis}
\end{tikzpicture}%
    \caption{\Ac{det} curves for different number of pilot symbols in the authentication phase, $K$.}
    \label{fig: num_pilot}
\end{figure}

\paragraph*{Number of Pilot Symbols in the Set-up Phase, $K_\mathrm{est}$}
\remembertext{Sensitivity}{We now consider the impact of the number of pilot symbols used for channel estimation in the set-up phase of CR-PLA. When $K_{\rm est}$ is finite, Bob adapts the threshold $\phi$ to match the desired \ac{fa} probability. In fact, for a finite number of reference gains estimation pilots $K_{\rm est}$, the threshold $\phi$ required to obtain a desired \ac{fa} probability is computed numerically, as well as the payoff matrix in \eqref{eq:pmd_matrix} (containing the \ac{md} probabilities) since the test distribution in \eqref{eq:LLTTest} cannot be computed in closed form.  We consider three scenarios: the perfect gain estimation ($K_{\rm est} = \infty$, our baseline), the Symmetric (Sym) scenario, and the Asymmetric (Asym) scenario. In the Sym case, Bob and Trudy compute the new strategies based on the new threshold and MD probability matrix. Finally, in the Asym case, Bob still keeps the baseline strategy, not taking into account the new setup (i.e., he deviates from the \ac{ne}), and Trudy, aware of this, designs her strategy to maximize the \ac{md} probability. 

Fig. \ref{fig: non_opt_stim} shows the obtained performance and we note that, depending on the desired \ac{fa} probability, the impact of this imperfection soon vanishes by increasing the number of pilots $K_{\rm est}$. We remark that such a high number of pilots is necessary only in the set-up phase, i.e., once and at the beginning of the protocol. In this simulation, we used a grid of $M_{\rm A}=M_{\rm T}=4\times4$ positions.}

\paragraph*{Number of Pilot Symbols for Channel Estimation $K$} About the impact of the number of pilot symbols used for channel estimation in the authentication phase of CR-PLA, Fig.~\ref{fig: num_pilot} shows the \ac{det} curves of the proposed solution for $N=1$ round and  $K=50$, $70$, and $100$ pilot symbols. We see that a higher $K$ yields better channel estimates and thus lower \ac{md} probabilities. However, the benefits are limited, indicating that it may be possible to achieve a good security performance even with low $K$ values, thus potentially saving drone energy and time. 

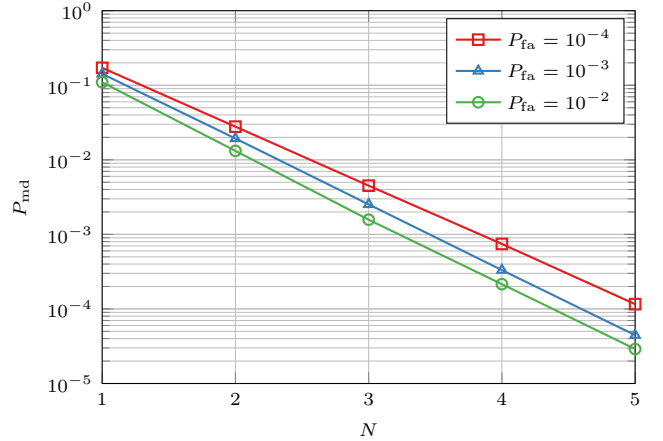
\begin{figure}
    \centering
%
%
\definecolor{mycolor1}{RGB}{228,26,28}%
\definecolor{mycolor2}{RGB}{55,126,184}%
\definecolor{mycolor3}{RGB}{77,175,74}%
\begin{tikzpicture}[ every plot/.style={thick}]

\begin{axis}[%
width=\fwidth,
height=0.7\fheight,
scale only axis,
xmin=1,
xmax=5,
xtick={1, 2, 3, 4, 5},
xlabel={$N$},
ymode=log,
ymin=1e-05,
ymax=1,
yminorticks=true,
ylabel={$P_{\rm md}$},
xmajorgrids,
ymajorgrids,
yminorgrids,
legend style={legend cell align=left, align=left}
]
\addplot [color=mycolor1, thick,mark=square, mark options={solid, mycolor1}]
  table[row sep=crcr]{%
1	0.1713855\\
2	0.0278555\\
3	0.004503\\
4	0.0007445\\
5	0.000116\\
};
\addlegendentry{$P_{\rm fa}= 10^{-4}$}

\addplot [color=mycolor2, thick,mark=triangle, mark options={solid, mycolor2}]
  table[row sep=crcr]{%
1	0.1422995\\
2	0.019327\\
3	0.0025265\\
4	0.000332\\
5	4.45e-05\\
};
\addlegendentry{$P_{\rm fa}= 10^{-3}$}

\addplot [color=mycolor3, thick,mark=o, mark options={solid, mycolor3}]
  table[row sep=crcr]{%
1	0.109667\\
2	0.013172\\
3	0.001574\\
4	0.000215\\
5	2.9e-05\\
};
\addlegendentry{$P_{\rm fa}= 10^{-2}$}

\end{axis}
\end{tikzpicture}%
    \caption{$P_{\rm md}$ as a function of the number of rounds, $N$, for $P_{\rm fa}= 10^{-2}$, $ 10^{-3}$, and  $10^{-4}$.}
    \label{fig: pmd_vs_rounds}
\end{figure}

\paragraph*{Number of Rounds $N$}  We now consider the impact of the number of rounds $N$ on the security performance. Fig.~\ref{fig: pmd_vs_rounds} shows the \ac{md} probability as a function of rounds $N$ for \ac{fa} probabilities $10^{-2}$, $10^{-3}$, and $10^{-4}$. From the results, we see that the \ac{md} decreases exponentially with the number of rounds, thus, we can easily reduce the \ac{md} probability to desired values by adding a few more rounds. This, however, comes at the cost of increased consumed energy, as discussed below.

\paragraph*{Two Drone Authentication}
\remembertext{multipleDrones}{We consider a scenario where Alice and Trudy maneuver two drones each. To avoid any possible collision, Alice set the drones to fly at different altitudes, $h_1$ and $h_2$. Additionally, Alice, to avoid introducing any degree of correlation that the attacker may exploit, flies her drones to move independently from one another. 
In turn, to maximize her chance of success, Trudy sets each of her own drones to fly independently but at the same altitudes of Alice's drones $h_1$ and $h_2$. As each Alice-Trudy pair of drones plays independently from the others, each game is solved using \eqref{eq:opt_problem}, and leveraging on the independence, we use test \eqref{eq:LLTTest} considering $N = 2$, as the shadowing maps and thus reference gains are known.
On the other hand, we remark that games played at different rounds are performed on the same shadowing map, while games played at the same time but at different altitudes are played on a different shadowing map. Thus, while in the $N=2$ round case, the same strategy can be used in both games, for the two drones (single round) scenario, a new strategy has to be computed for each drone pair. 

Fig. \ref{fig:pmd_pfa_swarms} shows the \ac{det} curves comparing the scenarios where Alice and Trudy control either one or two drones each, for $\sigma_{(s)_{\rm dB}}=6$ and \SI{10}{\decibel}. In particular, for the two-drones scenario, the drones are at altitudes $h_1 = \SI{30}{\meter}$ and  $h_2=\SI{70}{\meter}$.
Indeed, having two drones makes the test more robust,  approximately reducing the \ac{md} by a factor of $10$. %
Finally, we point out that the proposed approach may be straightforwardly extended to swarm authentication, placing the drones at different altitudes.} \remembertext{fig7comm}{We also report the results for both the urban and rural areas (in terms of shadowing). Even in this case, we note that a higher shadowing variance improves the security of the authentication mechanism.}

\begin{figure}
    \centering
%
%
\definecolor{mycolor1}{rgb}{0.00000,0.44700,0.74100}%
\definecolor{mycolor2}{rgb}{0.85000,0.32500,0.09800}%
\definecolor{mycolor3}{rgb}{0.92900,0.69400,0.12500}%
\definecolor{mycolor4}{rgb}{0.49400,0.18400,0.55600}%
\begin{tikzpicture}[ every plot/.style={thick}]

\begin{axis}[%
width=\fwidth,
height=0.7\fheight,
at={(0.758in,0.481in)},
scale only axis,
xmin=1e-3,
xmax=0.1,
xlabel style={font=\color{white!15!black}},
xlabel={$P_{\rm fa}$},
ymode=log,
ymin=0.005807,
ymax=0.157143615934803,
yminorticks=true,
ylabel style={font=\color{white!15!black}},
ylabel={$P_{\rm md}$},
axis background/.style={fill=white},
xmajorgrids,
ymajorgrids,
yminorgrids,
legend style={at={(0.98,0.5)}, anchor=east, legend cell align=left, align=left, draw=white!15!black},xlabel style={font=\scriptsize},ylabel style={font=\scriptsize},
]
\addplot [color=mycolor1, mark=diamond, mark options={solid, mycolor1}]
  table[row sep=crcr]{%
0.001	0.157143615934803\\
0.012	0.117402216589444\\
0.023	0.106499201643752\\
0.034	0.09936293753732\\
0.045	0.0940826064240942\\
0.056	0.089856101829295\\
0.067	0.0864106629098934\\
0.078	0.0830970046177835\\
0.089	0.0803498718946997\\
0.1	0.0778154243619187\\
};
\addlegendentry{$N_{\rm sw} = 1, \sigma_{(s)_{\rm dB}}=\SI{6}{\decibel}$}
\addplot [color=mycolor2, mark=o, mark options={solid, mycolor2}]
  table[row sep=crcr]{%
0.001	0.130828156575401\\
0.012	0.0996245669640356\\
0.023	0.090310387569679\\
0.034	0.084385002174113\\
0.045	0.0798572752468849\\
0.056	0.0763078936689745\\
0.067	0.0732453119523262\\
0.078	0.0706059574782674\\
0.089	0.0682434897430166\\
0.1	0.0660969103406652\\
};
\addlegendentry{$N_{\rm sw} = 1, \sigma_{(s)_{\rm dB}}= \SI{10}{\decibel}$}

\addplot [color=mycolor3, mark=square, mark options={solid, mycolor3}]
  table[row sep=crcr]{%
0.001	0.0262025\\
0.012	0.0154675\\
0.023	0.0132115\\
0.034	0.0117995\\
0.045	0.0107855\\
0.056	0.0100635\\
0.067	0.0094075\\
0.078	0.0089335\\
0.089	0.0083415\\
0.1	0.00806\\
};
\addlegendentry{$N_{\rm sw} = 2, \sigma_{(s)_{\rm dB}}= \SI{6}{\decibel}$}

\addplot [color=mycolor4, mark=triangle, mark options={solid, mycolor4}]
  table[row sep=crcr]{%
0.001	0.0187545\\
0.012	0.011422\\
0.023	0.009494\\
0.034	0.008525\\
0.045	0.0077045\\
0.056	0.0073305\\
0.067	0.0068835\\
0.078	0.006453\\
0.089	0.0060985\\
0.1	0.005807\\
};
\addlegendentry{$N_{\rm sw} = 2, \sigma_{(s)_{\rm dB}}= \SI{10}{\decibel}$}

\end{axis}

\end{tikzpicture}%
    \caption{\textcolor{black}{\ac{det} curves with $\sigma_{(s)_{\rm dB}}=\SI{6}{\decibel}$ and \SI{10}{\decibel}, for the single drones $N_{\rm sw}=1$ flying at $h_1=\SI{30}{\meter}$, and swarms $N_{\rm sw}=2$ scenario where an Alice-Trudy drone pair flies at $h_1 =\SI{30}{\meter} $ and the other pair flies at $h_2=\SI{70}{\meter}$}.} 
    \label{fig:pmd_pfa_swarms}
\end{figure}

\subsection{Impact of The Map Geometry} \label{sec:map_g_fin}

We now investigate the maps' effects on the proposed solution's performance. 

\paragraph*{Drone Altitude} First, we investigate the impact of Alice's and Trudy's altitude $h$. Fig.~\ref{fig:pmd_altezza} shows the achieved average $P_{\rm md}$ as a function of $h$, for $P_{\rm fa} = 10^{-2}$, $10^{-3}$, and $10^{-4}$, with $N=1$ round. We note that the \ac{md} probability slightly increases with $h$: indeed, increasing the altitude decreases the gain map diversity because the pathloss component becomes approximately constant as the grid area is fixed. In other words, the path loss becomes less relevant as all the grid positions experience the same one as the altitude $h$ increases. This leads to a reduced variance of the gains over the map, and thus more similar challenges/responses, which in turn makes it easier for Trudy to guess the correct response. 

\begin{figure}
    \centering
%
%
\definecolor{mycolor1}{RGB}{228,26,28}%
\definecolor{mycolor2}{RGB}{55,126,184}%
\definecolor{mycolor3}{RGB}{77,175,74}%
\begin{tikzpicture}[ every plot/.style={thick}]

\begin{axis}[%
width=0.94\fwidth,
height=0.7\fheight,
scale only axis,
y tick label style={/pgf/number format/.cd,fixed,precision=3},
xmin=10,
xmax=100,
xlabel={$h$ [\si{\meter}]},
ymin=0.08,
ymax=0.2,
ylabel={$P_{\rm md}$},
xmajorgrids,
ymajorgrids,
legend style={at={(0.02,0.98)}, anchor=north west, legend cell align=left, align=left}
]
\addplot [color=mycolor1,thick, mark=o, mark options={solid, mycolor1}]
  table[row sep=crcr]{%
10	0.135411376309334\\
20	0.144028946971675\\
30	0.152956180522737\\
50	0.161962583690137\\
70	0.170471804956518\\
100	0.18248825264795\\
};
\addlegendentry{$P_{\rm fa}=10^{-4}$}

\addplot [color=mycolor2, thick,mark=square, mark options={solid, mycolor2}]
  table[row sep=crcr]{%
10	0.113477632363345\\
20	0.120490356294335\\
30	0.12711342442743\\
50	0.134538745975567\\
70	0.141494021297209\\
100	0.152391342994134\\
};
\addlegendentry{$P_{\rm fa}=10^{-3}$}

\addplot [color=mycolor3, thick,mark=triangle, mark options={solid, mycolor3}]
  table[row sep=crcr]{%
10	0.0896981621595366\\
20	0.0948476093811435\\
30	0.0995068538807104\\
50	0.105204795272281\\
70	0.110906977892622\\
100	0.118864898214671\\
};
\addlegendentry{$P_{\rm fa}=10^{-2}$}

\end{axis}
\end{tikzpicture}%
    \caption{$P_{\rm md}$ as a function of the drone altitude, $h$, for $P_{\rm fa}= 10^{-3}$.}
    \label{fig:pmd_altezza}
\end{figure}
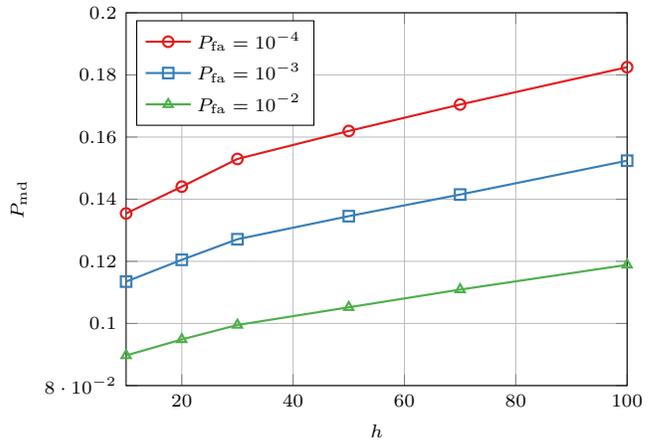

\begin{figure}
    \centering
%
%
\definecolor{mycolor1}{RGB}{228,26,28}%
\definecolor{mycolor2}{RGB}{55,126,184}%
\definecolor{mycolor3}{RGB}{77,175,74}%
\begin{tikzpicture}[ every plot/.style={thick}]

\begin{axis}[%
width=\fwidth,
height=0.7\fheight,
scale only axis,
xmin=0,
xmax=256,
xlabel={$M_{\rm A}$},
ymode=linear,
ymin=0.107224413744207,
ymax=0.357907359043289,
yminorticks=true,
ylabel={$P_{\rm md}$},
xmajorgrids,
ymajorgrids,
yminorgrids,
legend style={legend cell align=left, align=left}
]
\addplot [color=mycolor1,thick, mark=o, mark options={solid, mycolor1}]
  table[row sep=crcr]{%
4	0.357907359043289\\
16	0.241601665876758\\
36	0.212630993954295\\
64	0.198645759682681\\
100	0.185517186874759\\
144	0.177683394439551\\
196	0.168462864064971\\
256	0.166397889446334\\
};
\addlegendentry{$P_{\rm fa}= 10^{-4}$}

\addplot [color=mycolor2,thick, mark=square, mark options={solid, mycolor2}]
  table[row sep=crcr]{%
4	0.336038606464489\\
16	0.207199694086826\\
36	0.178302846661695\\
64	0.165586924727874\\
100	0.154125941893318\\
144	0.14736630928183\\
196	0.140505192907075\\
256	0.137932074869595\\
};
\addlegendentry{$P_{\rm fa}= 10^{-3}$}

\addplot [color=mycolor3,thick, mark=triangle, mark options={solid, mycolor3}]
  table[row sep=crcr]{%
4	0.311661126853534\\
16	0.171388730408268\\
36	0.14204209619491\\
64	0.130939078954272\\
100	0.121394276520396\\
144	0.114802054631668\\
196	0.109349715355722\\
256	0.107224413744207\\
};
\addlegendentry{$P_{\rm fa}= 10^{-2}$}

\end{axis}
\end{tikzpicture}%
    \caption{$P_{\rm md}$ as function of the number of grid positions, for $P_{\rm fa}= 10^{-2}$, $10^{-3}$, and $10^{-4}$.}
    \label{fig:density}
\end{figure}
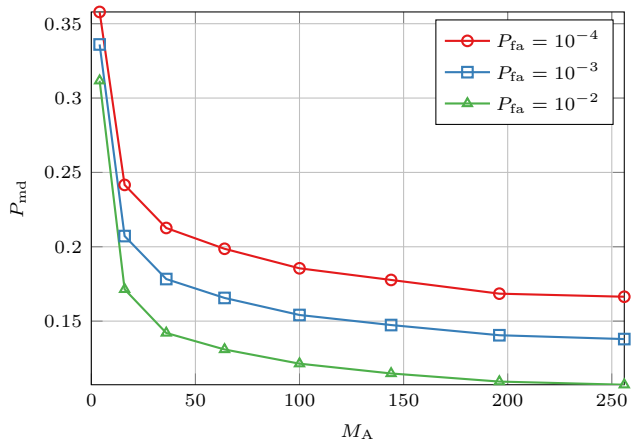
\paragraph*{Number of Positions on the Map} Then, we consider the impact of the number of positions on the map, with $\mathcal{X}=\mathcal{Y}$.
Fig.~\ref{fig:density} shows the \ac{md} probability as a function of the number of grid positions for $P_{\rm fa} = 10^{-2}$, $10^{-3}$, and $10^{-4}$, and $N=1$ round. We note that by increasing the overall number of positions (thus the number of possible challenges), the \ac{md} probability decreases. Still, after a rapid decrease, the \ac{md} reaches a plateau. Indeed, as the number of positions increases, the gains of nearby positions become closer, due to the shadowing correlation, thus not providing a significant further diversity for authentication. Still, the results highlight the trade-off between the duration of Step 1) of the protocol (see Section~\ref{sec:aut_scheme}) and the achievable performance, while showing that i) it is not necessary to sample too finely the space to achieve good performance and ii) a too precise sampling does not necessarily bring benefits for the legitimate party.

\begin{figure}
    \centering
%
%
\definecolor{mycolor1}{RGB}{228,26,28}
\definecolor{mycolor2}{RGB}{55,126,184}%
\definecolor{mycolor3}{RGB}{77,175,74}%
\definecolor{mycolor4}{RGB}{152,78,163}
\definecolor{mycolor5}{RGB}{251,154,153}
\begin{tikzpicture}[ every plot/.style={thick}]

\begin{axis}[%
width=\fwidth,
height=0.7\fheight,
scale only axis,
xmode=log,
xmin=0.001,
xmax=0.1,
xminorticks=true,
xlabel={$P_{\rm fa}$},
ymin=0,
ymax=0.25,
ylabel={$P_{\rm md}$},
xmajorgrids,
xminorgrids,
ymajorgrids,
legend style={legend cell align=left, align=left,only marks}
]
\addplot [color=mycolor1, thick,mark=halfcircle*, mark options={solid, mycolor3}]
  table[row sep=crcr]{%
0.001	0.178337415112495\\
0.00166810053720006	0.170764671784182\\
0.00278255940220713	0.162671207385019\\
0.00464158883361278	0.154439990903489\\
0.00774263682681127	0.146101587437535\\
0.0129154966501488	0.137357841499138\\
0.0215443469003188	0.128224247670952\\
0.0359381366380463	0.118534780612865\\
0.0599484250318941	0.107930050938522\\
0.1	0.0962736529590644\\
};
\addlegendentry{$\mathcal{X}=\mathcal{Y}$}

\addplot [color=mycolor1, thick,mark=square, mark options={solid, mycolor5}]
  table[row sep=crcr]{%
0.001	0.0673411223699494\\
0.00166810053720006	0.0623608546699659\\
0.00278255940220713	0.0571822266190865\\
0.00464158883361278	0.0519535579590643\\
0.00774263682681127	0.0465212402467019\\
0.0129154966501488	0.040732498495322\\
0.0215443469003188	0.0346688515191391\\
0.0359381366380463	0.0285582984519406\\
0.0599484250318941	0.0222964849424369\\
0.1	0.0163228458814048\\
};
\addlegendentry{$\bm{x}\neq\bm{y}$}

\addplot [color=mycolor1, thick,mark=triangle*, mark options={solid, mycolor4}]
  table[row sep=crcr]{%
0.001	0.216765299716727\\
0.00166810053720006	0.20803391116947\\
0.00278255940220713	0.198976768972438\\
0.00464158883361278	0.189356740708263\\
0.00774263682681127	0.179888564892078\\
0.0129154966501488	0.170021843894354\\
0.0215443469003188	0.159505454963863\\
0.0359381366380463	0.148425893931253\\
0.0599484250318941	0.136491028768164\\
0.1	0.123205426982707\\
};
\addlegendentry{$M_{\rm A}=M_{\rm T}/2$}

\addplot [color=mycolor2, thick,dashed, mark=halfcircle*, mark options={solid, mycolor3},forget plot]
  table[row sep=crcr]{%
0.001	0.165451917524245\\
0.00166810053720006	0.157877350627217\\
0.00278255940220713	0.15046613773407\\
0.00464158883361278	0.142762700654307\\
0.00774263682681127	0.134841805969725\\
0.0129154966501488	0.126543520503477\\
0.0215443469003188	0.117474209423507\\
0.0359381366380463	0.108276479984459\\
0.0599484250318941	0.0980524781055334\\
0.1	0.0868130422871972\\
};
\addlegendentry{Same, grid-points = 64}

\addplot [color=mycolor2, thick,dashed, mark=square, mark options={solid, mycolor5},forget plot]
  table[row sep=crcr]{%
0.001	0.0883466024774265\\
0.00166810053720006	0.0803513142131724\\
0.00278255940220713	0.0727076438812576\\
0.00464158883361278	0.0654287420210823\\
0.00774263682681127	0.0586572579593107\\
0.0129154966501488	0.052422871624009\\
0.0215443469003188	0.0465269742197313\\
0.0359381366380463	0.0406179729828203\\
0.0599484250318941	0.0345985589612311\\
0.1	0.0281923225667843\\
};
\addlegendentry{Not-same, grid-points = 64}

\addplot [color=mycolor2,thick, dashed, mark=triangle*, mark options={solid, mycolor4}, forget plot]
  table[row sep=crcr]{%
0.001	0.191777408989423\\
0.00166810053720006	0.183068637216819\\
0.00278255940220713	0.174752360373524\\
0.00464158883361278	0.165898821162634\\
0.00774263682681127	0.156902279615063\\
0.0129154966501488	0.147767309604458\\
0.0215443469003188	0.137221216832289\\
0.0359381366380463	0.126865732465457\\
0.0599484250318941	0.115372510805587\\
0.1	0.102849887181323\\
};
\addlegendentry{Half points, grid-points = 64}

\end{axis}
\end{tikzpicture}%
    \caption{\ac{det} curves with $M_{\rm T}=36$ (solid) and $M_{\rm T}=64$ (dashed) and various type of Alice-Trudy maps.}
    \label{fig: pmd_ma_ne_mt}
\end{figure}
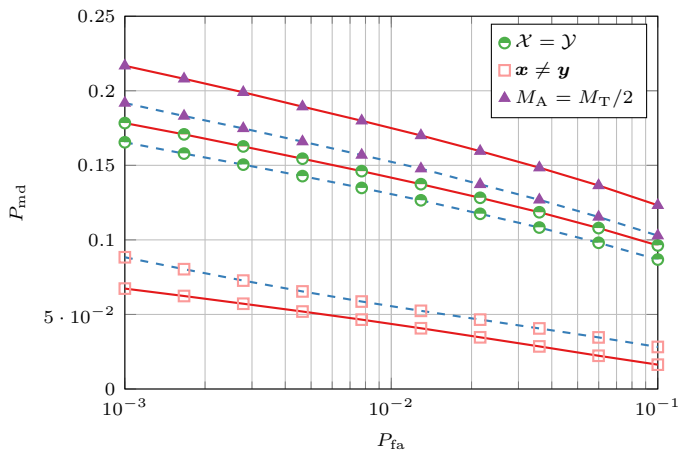

\paragraph*{Alice-Trudy Map Difference} We study the impact that different sets of positions available to Trudy and Alice have on the \ac{det} curves.  In particular, we consider the scenario (denoted as  $M_{\rm A} = M_{\rm T}/2$) wherein the positions in $\mathcal X$ are half of those in $\mathcal Y$ (thus $M_{\rm A} = M_{\rm T}/2$), and Alice's positions are selected uniformly at random from Trudy's map. We compare it with a scenario (denoted as $\mathcal X =\mathcal Y$) wherein Alice and Trudy have the same map. Lastly, we consider the case in which the maps of Alice and Trudy are the same, but Trudy is not allowed to select the same position as Alice (as this may lead to a clash), and denote this scenario as $\bm{x} \neq \bm{y}$.
Fig.~\ref{fig: pmd_ma_ne_mt} shows the \ac{det} for the three considered cases, comparing the results for $M_{\rm T} = 36$ and $64$, for $N=1$. We can see that when Alice's map has half of the positions of Trudy, the $P_{\rm md}$ increases, as expected. Still, there are minor differences even though the number of grid positions is halved for Alice: this suggests that as long as there is enough variability in the gains available, Bob can still achieve good security performance. This fact is also confirmed by the scenario where Trudy is deprived of {\em good} positions (i.e., where she's in the same Alice's location) and the \ac{md} decreases. These effects are more visible with fewer grid positions. Indeed, for a denser grid, it is less likely that the deleted positions are relevant.

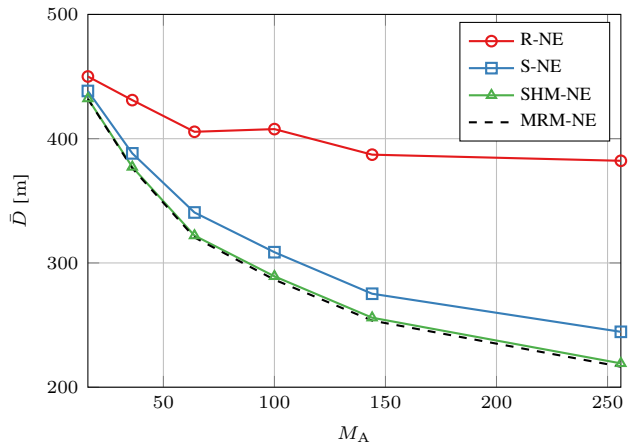
\begin{figure}
    \centering
%
%
\definecolor{mycolor1}{RGB}{228,26,28}%
\definecolor{mycolor2}{RGB}{55,126,184}%
\definecolor{mycolor3}{RGB}{77,175,74}%
\definecolor{mycolor4}{rgb}{0.49400,0.18400,0.55600}%
\begin{tikzpicture}

\begin{axis}[%
width=\fwidth,
height=0.7\fheight,
scale only axis,
xmin=16,
xmax=256,
xlabel={$M_{\rm A}$},
ymin=200,
ymax=500,
ylabel={$\bar{D}$ [\si{m}]},
xmajorgrids,
ymajorgrids,
legend style={legend cell align=left, align=left}
]
\addplot [color=mycolor1, thick,mark=o, mark options={solid, mycolor1}]
  table[row sep=crcr]{%
16	450.057219471115\\
36	431.032102846177\\
64	405.554411081948\\
100	407.711176171279\\
144	387.146752375477\\
256	382.206095060611\\
};
\addlegendentry{R-NE}

\addplot [color=mycolor2, mark=square,thick, mark options={solid, mycolor2}]
  table[row sep=crcr]{%
16	438.395559000929\\
36	388.192510638025\\
64	340.615564578194\\
100	308.650026381944\\
144	275.228842462536\\
256	244.582882233071\\
};
\addlegendentry{S-NE}

\addplot [color=mycolor3, mark=triangle,thick, mark options={solid, mycolor3}]
  table[row sep=crcr]{%
16	432.293220175672\\
36	377.221099255061\\
64	322.032256223522\\
100	289.231157238004\\
144	255.931966727777\\
256	219.197566134181\\
};
\addlegendentry{SHM-NE}

\addplot [color=black,thick, dashed]
  table[row sep=crcr]{%
16	431.973961057847\\
36	376.209942447775\\
64	320.549406657875\\
100	286.542177446994\\
144	253.606026871939\\
256	216.510719772194\\
};
\addlegendentry{MRM-NE}

\end{axis}
\end{tikzpicture}%
\caption{Average distance traveled by Alice vs the number of grid positions $M_{\rm A}$, with $N=2$ rounds, for $P_{\rm fa}= 10^{-3}$.}
    \label{fig: num_grid_points}
\end{figure}

\subsection{Average Traveled Distance}

We now discuss the effectiveness of the distance minimization algorithms. We consider the \ac{sne}, the optimal \ac{mrm}-\ac{ne}, and the \ac{shm}-\ac{ne} solutions. For comparison purposes, we also show the average distance for a \ac{ne} without distance optimization, a random \ac{ne} is selected at each round (R-NE). 

Fig.~\ref{fig: num_grid_points} shows the average distance $\bar{D}$ traveled by Alice over $N=2$ rounds, as a function of the number of positions of the map $M_{\rm A}$, positioned on a grid always occupying the same area. As we can see, the distance-optimized solutions outperform the baseline R-NE, and with a larger number of grid positions, the traveled distance decreases. This is because by increasing the positions the more degrees of freedom the optimization algorithms have, thus can find better \ac{ne}s from the distance perspective. We also notice that the \ac{shm}-\ac{ne} solution outperforms the \ac{sne} solution and achieves close-to-bound performance (represented by the optimal \ac{mrm}-\ac{ne} solution) but with much lower computations, confirming our suppositions.

\section{Conclusions}\label{sec:conclusion}
 
In this paper, we have proposed novel strategies for \ac{cr}-\ac{pla} with drones: we have optimized the \ac{pmd} used to select the challenge by Bob, and correspondingly, the \ac{pmd} of the random attack performed by Trudy. We formulated the problem as a zero-sum game with payoff the \ac{md} probability, thus characterizing the \acp{ne} of the game. To further decrease the \ac{md} probability, the protocol can be repeated for multiple rounds/movements.

Among solutions providing the same \ac{md} probability, we have then selected the \ac{pmd} for Bob, yielding the minimum energy consumption due to the movement. Three solutions were proposed: the optimal \ac{mrm}-\ac{ne} solution, the \ac{sne} solution where the challenge is chosen drawing always the same \ac{ne}, and the \ac{shm}-\ac{ne} solution where we associate the best \ac{ne} to each possible position on the map. \ac{shm}-\ac{ne} performs close to the optimal \ac{mrm}-\ac{ne} strategy, but it is much more lightweight in terms of computations.

Simulation results are obtained over realistic urban \ac{nlos} and rural scenarios, including path loss, fading, and shadowing phenomena. In particular, the proposed solution achieves \ac{md}/\ac{fa} probabilities of $\sim 10^{-3}$ with only $N=3$ rounds.

\bibliographystyle{IEEEtran}
\bibliography{IEEEabrv,biblio}

\begin{thebibliography}{10}
\providecommand{\url}[1]{#1}
\csname url@samestyle\endcsname
\providecommand{\newblock}{\relax}
\providecommand{\bibinfo}[2]{#2}
\providecommand{\BIBentrySTDinterwordspacing}{\spaceskip=0pt\relax}
\providecommand{\BIBentryALTinterwordstretchfactor}{4}
\providecommand{\BIBentryALTinterwordspacing}{\spaceskip=\fontdimen2\font plus
\BIBentryALTinterwordstretchfactor\fontdimen3\font minus \fontdimen4\font\relax}
\providecommand{\BIBforeignlanguage}[2]{{%
\expandafter\ifx\csname l@#1\endcsname\relax
\typeout{** WARNING: IEEEtran.bst: No hyphenation pattern has been}%
\typeout{** loaded for the language `#1'. Using the pattern for}%
\typeout{** the default language instead.}%
\else
\language=\csname l@#1\endcsname
\fi
#2}}
\providecommand{\BIBdecl}{\relax}
\BIBdecl

\bibitem{Adil23systematic}
M.~Adil, M.~A. Jan, Y.~Liu, H.~Abulkasim, A.~Farouk, and H.~Song, ``A systematic survey: {S}ecurity threats to {UAV}-aided {IoT} applications, taxonomy, current challenges and requirements with future research directions,'' \emph{IEEE Trans. Intell. Transp. Syst}, vol.~24, no.~2, pp. 1437--1455, Feb. 2023.

\bibitem{ceccato21}
M.~Ceccato, F.~Formaggio, and S.~Tomasin, ``Spatial {GNSS} spoofing against drone swarms with multiple antennas and {W}iener filter,'' \emph{IEEE Trans. Signal Process.}, vol.~68, pp. 5782--5794, Oct. 2020.

\bibitem{michieletto22}
G.~Michieletto, F.~Formaggio, A.~Cenedese, and S.~Tomasin, ``Robust localization for secure navigation of {UAV} formations under {GNSS} spoofing attack,'' \emph{IEEE Trans. Automat. Sci. and Eng}, pp. 1--14, Sept. 2022.

\bibitem{Illi2024Physical}
E.~Illi, M.~Qaraqe, S.~Althunibat, A.~Alhasanat, M.~Alsafasfeh, M.~de~Ree, G.~Mantas, J.~Rodriguez, W.~Aman, and S.~Al-Kuwari, ``Physical layer security for authentication, confidentiality, and malicious node detection: A paradigm shift in securing {IoT} networks,'' \emph{IEEE Commun. Surv. \& Tut.}, vol.~26, no.~1, pp. 347--388, Firstquarter 2024.

\bibitem{Hoang2024Physical}
T.~M. Hoang, A.~Vahid, H.~D. Tuan, and L.~Hanzo, ``Physical layer authentication and security design in the machine learning era,'' \emph{IEEE Commun. Surv. \& Tut.}, vol.~26, no.~3, pp. 1830--1860, Thirdquarter 2024.

\bibitem{Maeng2021Power}
S.~J. Maeng, Y.~Yapıcı, I.~Güvenç, H.~Dai, and A.~Bhuyan, ``Power allocation for fingerprint-based {PHY}-layer authentication with {mmWave} {UAV} networks,'' in \emph{Proc. of the Int. Conf. on Commun. (ICC)}, 2021, pp. 1--6.

\bibitem{Yazdinejad2021Federated}
A.~Yazdinejad, R.~M. Parizi, A.~Dehghantanha, and H.~Karimipour, ``Federated learning for drone authentication,'' \emph{Ad Hoc Netw.}, vol. 120, p. 102574, Sept. 2021.

\bibitem{gupta2014cryptography}
P.~C. Gupta, \emph{Cryptography and network security}.\hskip 1em plus 0.5em minus 0.4em\relax PHI Learning, 2014.

\bibitem{Tomasin22Challenge}
S.~Tomasin, H.~Zhang, A.~Chorti, and H.~V. Poor, ``Challenge-response physical layer authentication over partially controllable channels,'' \emph{IEEE Commun. Mag.}, vol.~60, no.~12, pp. 138--144, Dec. 2022.

\bibitem{Mazzo23Physical}
F.~Mazzo, S.~Tomasin, H.~Zhang, A.~Chorti, and H.~V. Poor, ``Physical-layer challenge-response authentication for drone networks,'' in \emph{Proc. of IEEE Global Commun. Conf. (GLOBECOM)}, 2023, pp. 3282--3287.

\bibitem{ardizzon2024energy}
\BIBentryALTinterwordspacing
F.~Ardizzon, D.~Salvaterra, M.~Piana, and S.~Tomasin, ``Energy-based optimization of physical-layer challenge-response authentication with drones,'' [Accepted] Proc. of IEEE Global Commun. Conf. (GLOBECOM), 2024. [Online]. Available: \url{https://arxiv.org/abs/2405.03608}
\BIBentrySTDinterwordspacing

\bibitem{Abeywickrama18}
H.~V. Abeywickrama, B.~A. Jayawickrama, Y.~He, and E.~Dutkiewicz, ``Comprehensive energy consumption model for unmanned aerial vehicles, based on empirical studies of battery performance,'' \emph{IEEE Access}, vol.~6, pp. 58\,383--58\,394, Oct. 2018.

\bibitem{Zhang2020Lightweight}
Y.~Zhang, D.~He, L.~Li, and B.~Chen, ``A lightweight authentication and key agreement scheme for {I}nternet of {D}rones,'' \emph{Comput. Commun.}, vol. 154, pp. 455--464, Mar. 2020.

\bibitem{Xiao2016Channel}
L.~Xiao, T.~Chen, G.~Han, W.~Zhuang, and L.~Sun, ``Channel-based authentication game in {MIMO} systems,'' in \emph{Proc. of IEEE Global Commun. Conf. (GLOBECOM)}, 2016, pp. 1--6.

\bibitem{Ge2024GAZETA}
Y.~Ge and Q.~Zhu, ``{GAZETA}: Game-theoretic zero-trust authentication for defense against lateral movement in {5G} {IoT} networks,'' \emph{IEEE Trans. on Inf. Forens. and Secur.}, vol.~19, pp. 540--554, Oct. 2024.

\bibitem{Wu2023Game}
Y.~Wu, T.~Jing, Q.~Gao, Y.~Wu, and Y.~Huo, ``Game-theoretic physical layer authentication for spoofing detection in {I}nternet of {T}hings,'' \emph{Digit. Commun. and Netw.}, Jan. 2023.

\bibitem{Zhou2022Game}
Y.~Zhou, P.~L. Yeoh, K.~J. Kim, Z.~Ma, Y.~Li, and B.~Vucetic, ``Game theoretic physical layer authentication for spoofing detection in {UAV} communications,'' \emph{IEEE Trans. on Veh. Technol.}, vol.~71, no.~6, pp. 6750--6755, Mar. 2022.

\bibitem{gudmundson1991correlation}
M.~Gudmundson, ``Correlation model for shadow fading in mobile radio systems,'' \emph{Electron. Lett.}, vol.~23, no.~27, pp. 2145--2146, Nov. 1991.

\bibitem{bentom}
N.~Benvenuto, G.~Cherubini, and S.~Tomasin, \emph{Algorithms for Commun. Systems and their Applications}, 2nd~ed.\hskip 1em plus 0.5em minus 0.4em\relax Wiley, 2021.

\bibitem{tse2005fundamentals}
D.~Tse and P.~Viswanath, \emph{Fundamentals of wireless communication}.\hskip 1em plus 0.5em minus 0.4em\relax Cambridge University Press, 2005.

\bibitem{Clarke1968statistical}
R.~H. Clarke, ``A statistical theory of mobile-radio reception,'' \emph{The Bell Syst. Tech. J.}, vol.~47, no.~6, pp. 957--1000, July-Aug. 1968.

\bibitem{Senigagliesi21Comparison}
L.~Senigagliesi, M.~Baldi, and E.~Gambi, ``Comparison of statistical and machine learning techniques for physical layer authentication,'' \emph{IEEE Trans. Inf. Forensics Secur.}, vol.~16, pp. 1506--1521, Oct. 2021.

\bibitem{Bragagnolo21Authentication}
L.~Bragagnolo, F.~Ardizzon, N.~Laurenti, P.~Casari, R.~Diamant, and S.~Tomasin, ``Authentication of underwater acoustic transmissions via machine learning techniques,'' in \emph{Proc. of IEEE Int. Conf. on Microw., Antennas, Commun. and Electron. Syst. (COMCAS)}, 2021, pp. 255--260.

\bibitem{Wyner1975wire-tap}
A.~D. Wyner, ``The wire-tap channel,'' \emph{The Bell System Technical Journal}, vol.~54, no.~8, pp. 1355--1387, 1975.

\bibitem{9409835}
I.~Bisio, C.~Garibotto, H.~Haleem, F.~Lavagetto, and A.~Sciarrone, ``On the localization of wireless targets: A drone surveillance perspective,'' \emph{IEEE Network}, vol.~35, no.~5, pp. 249--255, 2021.

\bibitem{Srigrarom2021Multi-camera}
S.~Srigrarom, N.~J.~L. Sie, H.~Cheng, K.~H. Chew, M.~Lee, and P.~Ratsamee, ``Multi-camera multi-drone detection, tracking and localization with trajectory-based re-identification,'' in \emph{Proc. 2021 Second International Symposium on Instrumentation, Control, Artificial Intelligence, and Robotics (ICA-SYMP)}, 2021, pp. 1--6.

\bibitem{Chiarello21Jamming}
L.~Chiarello, P.~Baracca, K.~Upadhya, S.~R. Khosravirad, and T.~Wild, ``Jamming detection with subcarrier blanking for {5G} and beyond in industry 4.0 scenarios,'' in \emph{Proc. Annu. Int. Symp. on Pers., Indoor and Mobile Radio Commun. (PIMRC)}, 2021, pp. 758--764.

\bibitem{Peng24GLRT}
X.~Peng, C.~Huang, X.~Zhu, Z.~Chen, and X.~Yuan, ``{GLRT}-based spacetime detection algorithms via joint {DoA} and {D}oppler shift method for {GNSS} spoofing interference,'' \emph{IEEE Internet Things J.}, pp. 1--1, June 2024.

\bibitem{ardizzon2024learning}
\BIBentryALTinterwordspacing
F.~Ardizzon and S.~Tomasin, ``Learning the likelihood test with one-class classifiers for physical layer authentication,'' 2024. [Online]. Available: \url{https://arxiv.org/abs/2210.12494}
\BIBentrySTDinterwordspacing

\bibitem{tadelis2013game}
S.~Tadelis, \emph{Game theory: an introduction}.\hskip 1em plus 0.5em minus 0.4em\relax Princeton University Press, 2013.

\bibitem{v1928theorie}
J.~V.~Neumann, ``{Zur Theorie der Gesellschaftsspiele},'' \emph{Math. Ann.}, vol. 100, no.~1, pp. 295--320, Dec. 1928.

\bibitem{washburn2014two}
A.~Washburn, \emph{Two-person zero-sum games}.\hskip 1em plus 0.5em minus 0.4em\relax Springer, 2014.

\bibitem{boggs1995sequential}
P.~T. Boggs and J.~W. Tolle, ``Sequential quadratic programming,'' \emph{Acta Numer.}, vol.~4, p. 1–51, 1995.

\bibitem{sharma2018study}
N.~Sharma, M.~Magarini, L.~Dossi, L.~Reggiani, and R.~Nebuloni, ``A study of channel model parameters for aerial base stations at 2.4 {GHz} in different environments,'' in \emph{Proc. IEEE Annu. Consumer Commun. \& Netw. Conf. (CCNC)}, 2018, pp. 1--6.

\bibitem{chandrasekharan2015propagation}
S.~Chandrasekharan, A.~Al-Hourani, K.~Magowe, L.~Reynaud, and S.~Kandeepan, ``Propagation measurements for {D2D} in rural areas,'' in \emph{Proc. of the Int. Conf. on Commun. Workshop (ICCW)}.\hskip 1em plus 0.5em minus 0.4em\relax IEEE, 2015, pp. 639--645.

\end{thebibliography}

\appendices
\section{Proof of Lemma~\ref{lemma:channelStat}} \label{app:lemma}
From \eqref{eq:filtered_sig} with $\bm{q}=\bm{x}(i)$, let  
\begin{equation} \label{eq:r_tilda}
    \tilde{r}_k= \tilde{g}_{\bm{x}(i)}h_ke^{-j\phi}+w'_k \,.
\end{equation}
We note that the real and imaginary parts of $\tilde{r}_k$ are zero-mean Gaussian distributed random variables with variance
\begin{equation} \label{eq:sigma_est}
    \tilde{\sigma}^2_{\bm{x}(i)}= \frac{\tilde{g}_{\bm{x}(i)}^2 +\sigma_w^2}{2}\,,
\end{equation}
which depends on Alice's position $\bm{x}(i)$.
Consequently, $\left| \tilde{r}_k\right|^2$ is exponentially distributed with parameter
\begin{equation} \label{eq:lda_estim}
    \lambda_{\bm{x}(i)} = \frac{1}{2\tilde{\sigma}^2_{\bm{x}(i)}}. 
\end{equation}
We assume that the samples $r_k$ of the received signal are taken sufficiently far apart in time, i.e., with $T_\mathrm{s}>T_\mathrm{c}$, so the fading coefficients $h_k$ are statistically independent. This implies that
\begin{equation} \label{eq:ind_2}
    \mathbb{E}[\tilde{r}_k\tilde{r}_{k'}^*]=0 \,, \forall k' \neq k\, ,
 \end{equation}
 thus samples taken at different times $k$ and $k'$ are independent, since $\tilde{r}_k$ is Gaussian. Hence, we resort to the central limit theorem, thus $\mu(\bm{x}(i))$, defined in \eqref{eq:filtered_sig}, converges, for $K$ sufficiently large, in terms of \ac{cdf} to a normal distribution 
\begin{equation}
    \sqrt{K}\left(\mu(\bm{x}(i))-\frac{1}{\lambda_{\bm{x}(i)}}\right) \xrightarrow{d} \mathcal{N}\left(0, \frac{1}{\lambda_{\bm{x}(i)}^2}\right)\,.
\end{equation}
Lastly, reordering the terms and defining  $m(i)={1}/{\lambda_{\bm{x}(i)}}$, we obtain \eqref{eq:central_limit_Alice}.

\section{Proof of Lemma~\ref{lemma:ind_Alice}}\label{app:ind_Alice}
From \eqref{eq:filtered_sig} and \eqref{eq:r_tilda}, two gains $\hat{m}_n$ and $\hat{m}_{n'}$ estimated by Bob at different times $n=kT_s$ and $n'=k'T_s$ have correlation
    \begin{equation} 
        \begin{split}
            \mathbb{E}[\hat{m}_{n}\hat{m}_{n'}] &= \mathbb{E}\left[\frac{1}{K} \sum_{k}\left| \tilde{r}_k \right|^2 \frac{1}{K} \sum_{k'}\left| \tilde{r}_{k'} \right|^2\right] \\
            &= \frac{1}{K^2} \sum_{k} \sum_{k'} \mathbb{E}\left[|\tilde{r}_k|^2|\tilde{r}_{k'}|^2\right].
        \end{split}
    \end{equation}
Next, as discussed in Appendix~\ref{app:lemma}, samples $\tilde{r}_k$ and $\tilde{r}_{k'}$ measured at sufficiently far apart instants, and therefore also their squared amplitude, are independent, hence reordering the terms
\begin{equation}\begin{split}\label{eq:ind}
    \mathbb{E}[\hat{m}_{n}\hat{m}_{n'}] & =  \frac{1}{K^2} \sum_{k} \sum_{k'} \mathbb{E}\left[|\tilde{r}_k|^2|\tilde{r}_{k'}|^2\right] \\ 
   & = \frac{1}{K^2} \sum_{k} \sum_{k'} \mathbb{E}\left[|\tilde{r}_k|^2\right] \mathbb{E}\left[ |\tilde{r}_{k'}|^2\right]  = \mathbb{E}[\hat{m}_{n}]\mathbb{E}[\hat{m}_{n'}]. 
 \end{split}\end{equation}
From \eqref{eq:ind}, we have that $\hat{m}_{n}$ and $\hat{m}_{n'}$ are uncorrelated, thus independent as they are Gaussian.

\section{Proof of Theorem~\ref{th:trudyDinamic}}\label{app:ind_trudy}
As discussed in Section~\ref{at_sec}, Bob draws the gains independently, thus \eqref{eq:alice_pa} holds. Given that Bob's and Trudy's strategies are independent of each other, the \ac{md} probability can be written as
\begin{equation}\label{eq:ind1}
\begin{split}
    P_{\rm md} =&\sum_{\bm m}\sum_{\bm m_{\rm T}} {\mathbb P}\left(\sum_{n=1}^N  \frac{K (\hat{m}_n(\bm{m}_{\rm T})-m_n)^2}{m_n^2} \leq \varphi \Big|\bm{m},\bm{m}_{\rm T} \right) \times \\ 
       &p_{\rm A}(\bm{m})p_{\rm T}(\bm{m}_{\rm T})\,,
    \end{split}
\end{equation}
 where \eqref{eq:ind1} is \eqref{eq:pmd_complete} highlighting the fact that, in general, the measured gain when Trudy is transmitting may be a function of the previous (or next) gains, i.e., $z_n =z_n(\bm{m}_{\rm T})=\hat{m}_n(\bm{m}_{\rm T})$.
 
Focusing on the choice of $m_{{\rm T},k}$, $k=1, \ldots, N$, let $\bar{\bm{m}}_{\rm T}$ be the vector $\bm{m}_{\rm T}$ from which we exclude the $k$-th entry $m_{{\rm T},k}$. Next, we define
\begin{equation}\label{eq:auxVar}
    \phi(\bar{\bm{m}}_{\rm T}, \bm{m}) = \sum_{n=1, n\neq k}^N  \frac{K (\hat{m}_n(m_{{\rm T},n})-m_n)^2}{m_n^2}\,,
\end{equation}
representing the results of all the tests except the $k$-th test.

To prove the theorem, we focus on the latter term in \eqref{eq:ind1}. In particular, we split the $k$-th test result from the others exploiting \eqref{eq:auxVar}, as 
\begin{equation}  
\begin{split}
     & {\mathbb P}\left(\sum_{n=1}^N  \frac{K (\hat{m}_n(\bm{m}_{\rm T})-m_n)^2}{m_n^2} \leq \varphi\Big|\bm{m},\bm{m}_{\rm T} \right) = \\
     & {\mathbb P}(\phi(\bar{\bm{m}}_{\rm T}, \bm{m}) = \phi|\bar{\bm{m}}_{\rm T}, \bm{m} ) \times \\ 
     & {\mathbb P}\left( \frac{K (\hat{m}_k(m_{{\rm T},k})-m_k)^2}{m_k^2} \leq \varphi - \phi\Big|m_k, m_{{\rm T},k}, \phi \right)
\end{split}
\end{equation}

Now, noticed that in \eqref{eq:ind1} it holds 
\begin{equation}
    p_{\rm T}(\bm{m}_{\rm T}) = {\mathbb P}(m_{{\rm T},k}|\bar{\bm{m}}_{\rm T}){\mathbb P}(\bar{\bm{m}}_{\rm T})\,,
\end{equation}
by reordering the terms in \eqref{eq:ind1}, we can define the probability
\begin{equation}\begin{split}\label{eq:probSingle}   
f(\bm{m}_{\rm T}, \bm{m}) &=  {\mathbb P}(m_{{\rm T},k}|\bar{\bm{m}}_{\rm T}) \times \\ 
 &{\mathbb P}\left( \frac{K (\hat{m}_k(m_{{\rm T},k})-m_k)^2}{m_k^2} \leq \varphi - \phi\Big|m_k, m_{{\rm T},k}, \phi \right), 
\end{split}
\end{equation}
which groups together all the terms that are impacted by the choice of the gain $ m_{{\rm T},k}$ at round $k$ in \eqref{eq:ind1}. We note that $m_n$, $n=1, \ldots, N$, are independent and that the maximization of this probability depends on neither $\varphi$ nor $\phi$. 
In other words, this shows that on each round Trudy aims at maximizing \eqref{eq:probSingle}, regardless of the previous and next choices. 

Finally, we can conclude that each round is an independent game that can be solved separately, i.e., the overall attacker \ac{pmd} can be factorized as in \eqref{eq:attackFactoriz}.
 
\section{Proof of Lemma~\ref{lemma:same_payoff}}\label{app:same_payoff}
Let $(\bm{a}_1^\star,\bm{e}_1^\star)$ and $(\bm{a}_2^\star,\bm{e}_2^\star)$ be two \ac{ne}s. Then, by definition of \ac{ne}, we have that
\begin{equation} \label{eq:app_1}
\begin{split}
    u_{\rm A}(\bm{a}_1^\star,\bm{e}_1^\star) \geq u_{\rm A}(\bm{a}_2^\star,\bm{e}_1^\star), \\
    u_{\rm T}(\bm{a}_1^\star,\bm{e}_1^\star) \geq u_{\rm T}(\bm{a}_1^\star,\bm{e}_2^\star).
\end{split}
\end{equation}
Since the game is zero-sum, we have $u_{\rm T}=-u_{\rm A}$, thus $u_{\rm A}(\bm{a}_2^\star,\bm{e}_1^\star) \geq u_{\rm A}(\bm{a}_2^\star,\bm{e}_2^\star)$. Hence 
\begin{equation} \label{eq:app_2}
    u_{\rm A}(\bm{a}_1^\star,\bm{e}_1^\star) \geq u_{\rm A}(\bm{a}_2^\star,\bm{e}_1^\star)  \geq u_{\rm A}(\bm{a}_2^\star,\bm{e}_2^\star)\,,
\end{equation}
and similarly
\begin{equation} \label{eq:app_3}
    u_{\rm A}(\bm{a}_1^\star,\bm{e}_1^\star) \leq u_{\rm A}(\bm{a}_1^\star,\bm{e}_2^\star)  \leq u_{\rm A}(\bm{a}_1^\star,\bm{e}_2^\star)\,.
\end{equation}
Following the inequalities of \eqref{eq:app_2} and \eqref{eq:app_3} we have that
\begin{equation} \label{eq:app_4}
    u_{\rm A}(\bm{a}_1^\star,\bm{e}_1^\star) = u_{\rm A}(\bm{a}_2^\star,\bm{e}_2^\star)\,,
\end{equation}
i.e., the two \ac{ne}s yield the same utility.

\clearpage

\onecolumn

\end{document}